\documentclass[prb,pre,aps,twocolumn,amsmath,amssymb,floatfix,
superscriptaddress]{revtex4-1}
\usepackage[dvips]{graphics}
\usepackage{graphicx}
\usepackage{color}
\usepackage{blindtext}
\usepackage{amsmath}
\definecolor{dred}{rgb}{0.75,0,0}
\usepackage{graphicx}
\usepackage{dcolumn}
\usepackage{bm}
\usepackage{xcolor}
\usepackage{braket}
\usepackage{physics}
\usepackage{appendix}
\usepackage{hyperref}
\hypersetup{
 colorlinks=true,
citecolor=magenta,filecolor=blue,
 linkcolor=magenta,
 }

\usepackage{graphicx}
\usepackage{dcolumn}
\usepackage{bm}
\usepackage{xcolor}

\definecolor{codegreen}{rgb}{0,0.6,0}
\definecolor{codegray}{rgb}{0.5,0.5,0.5}
\definecolor{codepurple}{rgb}{0.58,0,0.82}
\definecolor{backcolour}{rgb}{0.95,0.95,0.92}
 
\begin{document}

\preprint{APS/123-QED}

\title{\textcolor{blue}{Engineering complete delocalization of single particle states in a class of one dimensional aperiodic lattices: a quantum dynamical study}} 

\author{Sougata Biswas}
\affiliation{Department of Physics, Presidency University, 86/1 College Street, Kolkata, West Bengal - 700 073, India}
\affiliation{sougata.rs@presiuniv.ac.in}
\author{Arunava Chakrabarti}
\affiliation{Department of Physics, Presidency University, 86/1 College Street, Kolkata, West Bengal - 700 073, India}
\affiliation{arunava.physics@presiuniv.ac.in}
\date{\today}

\begin{abstract}
We study quantum dynamics of a wave packet on a class of one dimensional decorated aperiodic lattices, described within a tight binding formalism. We look for the possibility of finding extended single particle states even in the absence of any translational periodicity. The chosen lattices are stubbed with one or more atoms, tunnel coupled to the backbone, thereby introducing a minimal quasi-one dimensionality. It is seen that, for a group of such lattices a certain correlation between the {\it numerical values} of the hopping amplitudes leads to a {\it complete delocalization} of single particle states. In some other cases, a special value of a magnetic flux trapped in the loops present in the geometries delocalize the states, leading to a {\it flux driven} insulator to metal transition. The mean square displacement, temporal autocorrelation function, the time dependence of the inverse participation ratio, or the information entropy - the so-called hallmarks of studying localization based on dynamics - all of them indicate such a complete turnover in the nature of the single particle states and the character of the energy spectrum under suitable conditions. The results shown in this work using quasiperiodic lattices of the Fibonacci family are much more general and hold good even for a randomly disordered arrangement of the building blocks of the systems considered, and indicate a subtle universality class under which these lattices can be grouped. 
\end{abstract}

\maketitle

\section{Introduction}

The study of disorder driven localization of waves, pioneered by Anderson~\cite{anderson}, and followed up by a series of results ~\cite{lee,abrahams,borland} and reviews~\cite{kramer,lagendijk}, has remained an area of everlasting research interest. The general, broad understanding is that, a random, uncorrelated disorder will inhibit wave propagation, resulting in a spatial confinement of the wave, `localizing' it on a finite domain of the underlying lattice. The effect is strongest in one dimension where (almost) all the single states are exponentially localized no matter how weak the disorder is~\cite{borland}. The effect of disorder is similar in two dimensions~\cite{abrahams}, while in three dimensions there is the possibility of observing an insulator-metal transition across a {\it mobility edge}~\cite{mott}. A series of experiments involving light~\cite{albada,weirsma,martin,segev}, matter waves and ultracold atoms~\cite{roati,white} have established the fact and have invigorated research in this area in recent times. 

A series of results evolved in the last three decades, drawing attention to a whole lot of non-trivial variations of the phenomenon of localization. This area of research discussed the existence of {\it extended}, i.e., unscattered eigenstates in certain classes of one dimensional lattices that lacked translational invariance. The beginning of this era can be attributed to the pioneering work on the so called `random dimer model' (RDM)~\cite{dunlap,phillips}, where a local {\it positional correlation} between the constituent `atoms' was shown to lead to the unscattered eigenfunctions in an otherwise disordered one dimensional binary alloy model. The disorder, in the RDM, has a `structure' in it, and is {\it positionally correlated}~\cite{dunlap,phillips}. The outcome of having a such kind of disorder was experimentally verified through a measurement of the conductivity of positionally correlated multilayered structures~\cite{bellani}, and more recently, by studying the anomalous wave packet evolution in an optical waveguide~\cite{alex}. The RDM was later generalized to the case of random $N$-mer systems~\cite{wu,izrailev,kosior,major}, as well as to certain long-range correlated tight binding lattice models where numerical search revealed patches of continua, populated by extended eigenstates, and a possible insulator-metal transition, and even mobility edges~\cite{moura,krokhin}.


\begin{figure}[ht]
\centering
\includegraphics[width=\columnwidth]{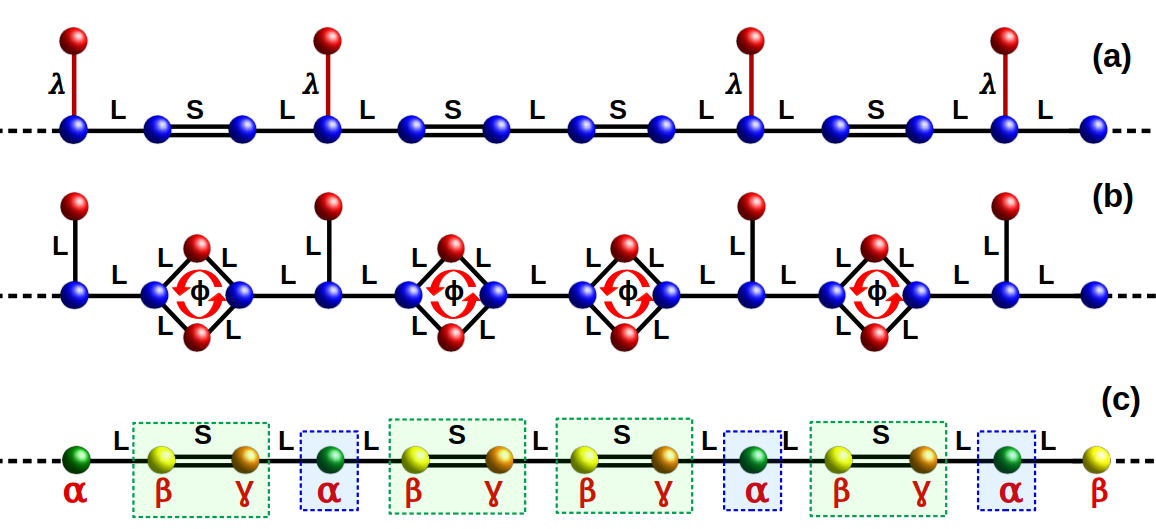}

\caption{(Color online) The Fibonacci array of (a) `dimer-stub' and (b) `diamond-stub' quantum networks, where each diamond plaquette is threaded by a constant magnetic flux $\Phi$. (c) The renormalized version of (a) and (b) is obtained after decimating the red-colored sites.}  
\label{fig}
\end{figure}

Further studies revealed that a whole lot of quasiperiodic and fractal lattices that are non-translationally invariant but deterministic, and hierarchically grown systems, can support an infinite number of extended non-Bloch states as well~\cite{enrique,ac1,ac2,ac3}. These latter results are different from the basic results of the RDM and its analogues in the sense that, the hierarchical pattern of growth allows any local positional correlation to be found at {\it every scale of length}. This leads to the existence of an infinite number of extended states for the quasiperiodic or the fractal lattices in the thermodynamic limit. But, the extended states in all these cases are found to be separated from each other, and the appearance of continuous, gapless `bands' of extended states, though strongly suggested by exhaustive numerical studies, still needs to be explored in detail.

The results mentioned above, all point to the existence of extended states in non-translationally invariant systems, though the localized eigenfunctions still persist as an obvious consequence of disorder. The spectra, in all the systems described above, are mixed ones.
The question that we address ourselves in the present communication is, can we really have a disordered system, especially in one or quasi-one dimension, where {\it all} states should be extended, and the energy spectrum will be {\it absolutely continuous}? The answer, seemingly, is `yes'. It happens for a wide class of {\it decorated} lattices. The `decoration' is done by attaching atomic sites to the principal one dimensional chain, namely, the {\it backbone}, at selective lattice points. In some cases, one can even think of a magnetic flux trapped in the `cells' (some structures with loops) that build up the chain. A trapped magnetic flux is already known to result in very interesting behaviour in the magnetoresistance of mesoscopic rings~\cite{amato}, and we intend to examine the effect of magnetic flux in the present case also.

It has recently been reported ~\cite{biplab1,biplab2,atanu,amrita} that, one dimensional lattices, described by a tight binding Hamiltonian with only
an off-diagonal (hopping) disorder can indeed give rise to an absolutely continuous spectrum spanning the {\it entire range} of the allowed energy eigenvalues. This requires a definite correlation among the numerical values of the nearest neighbor `hopping integrals'.  This is a kind of a {\it correlation driven} insulator-metal transition, but unlike the RDM, the correlation is now between the {\it numerical values} of a subgroup of the system parameters, viz. the hopping (overlap) integrals. For systems where there is a magnetic flux $\Phi$ trapped in a subset of the unit cells, an additional choice of $\Phi=\Phi_0/4$ ($\Phi_0=hc/e$ is the fundamental flux quantum), can change an insulating system to a metallic one.  A very recent report of such correlation driven continuum in the energy spectrum of a fractal Koch curve consolidates these results~\cite{sougata}, where it is shown that the  {\it entire} energy spectrum of a Koch curve sheds off its fragmented nature, and turns out to be continuous, and the density of states resembles that of a perfectly periodic one dimensional lattice. All the eigenstates become extended in character. The end-to-end transport is practically unattenuated, and can achieve a high value, close to unity, for a long system. 

The previous works, or most of them, mentioned so far, were based on the solution of the time independent Schr\"{o}dinger equation. It is equally important to know the time dependent dynamics of a wave packet released on such {\it decorated} one or quasi-one dimensional lattices. In particular, the effect of the special correlation talked about above - the role of the relative magnitudes of the hopping integrals and the magnetic flux (where needed) on the long time behavior of the mean square displacement (MSD), the temporal auto correlation function (TAF) or the information entropy for such cases are yet to be studied in details, and this is our basic motivation.

We undertake an in-depth study of the quantum dynamics~\cite{katsanos,lozano,pastawski} of a wave packet on a class of lattices that are essentially one dimensional in character with one or more atoms side-coupled to the backbone. 

Two such examples, based on a quasiperiodic Fibonacci lattice are discussed in details in this paper, and are shown in Fig.~\ref{fig} (a) and (b). The analysis presented in the paper of course remains valid for any arrangement (placement) of the unit cells that build the lattice, be it quasiperiodic of any kind, or randomly disordered. The atoms, tunnel-coupled from the side are shown in red color. We work within a tight binding formalism and specifically intend to investigate the dynamics of the wave packet when the hopping (overlap) integrals in the Hamiltonian are numerically correlated in such a way that these non-translationally invariant systems have a continuous distribution of energy eigenvalue, and the wavefunctions turn out to be extended in character. Apart from a very recent work in this direction~\cite{arka}, there are no studies related to such decorated lattices, to the best of our knowledge, and we need to substantiate the ideas more rigorously.

To facilitate the understanding regarding the fundamental issues, we present, in Section II, a short coverage of the central results in terms of a time independent Schr\"{o}dinger equation describing our systems. The study of the dynamics is detailed out in Section III, and in Section IV we draw our conclusions.

\label{intro}

\section{The tight binding model and the basics}
The tight binding Hamiltonian in Wannier basis, is written as, 
\begin{equation} 
H = \epsilon \ket{n}\bra{n} + \sum_{<nm>} V_{nm} \ket{n}\bra{m} + h.c.
\label{hamiltonian}
\end{equation}
Here, $\epsilon$ is the uniform on-site potential (set equal to zero in all our numerical calculations), and $V_{nm}$ symbolizes the hopping amplitude between the nearest neighbors.

The first system we talk about is an array of {\it dimers} (the blue diatomic molecule) and {\it stubs} (a blue-red pair) arranged in a quasiperiodic Fibonacci sequence. A segment of an infinite dimer-stub array is shown in  Fig.~\ref{fig}(a). The placement of the dimers and the stubs follows the recursive growth pattern of a Fibonacci array, namely, $\mathcal{A} \rightarrow \mathcal{A} \mathcal{B}$ and $\mathcal{B} \rightarrow \mathcal{A}$, beginning, say, with the letter $\mathcal{A}$~\cite{kohmoto}. In Fig.~\ref{fig}(a) The paired blue circles stand for $\mathcal{A}$, and the vertical blue-red pair represents $\mathcal{B}$.

In our second example, as shown in Fig.~\ref{fig}(b), we have a Fibonacci sequence of diamond-shaped blocks and stubs (as before).
Now, the diamond shape stands for the letter $\mathcal{A}$ and the blue-red vertical stub, just as before, represents the letter $\mathcal{B}$. A constant magnetic flux $\Phi$ is trapped in each diamond plaquette. 

 For the `dimer-stub' Fibonacci sequence $V_{nm}=L$ between the base of a stub and a vertex (left or right) of a dimer. $V_{nm}=S$ between the atomic sites constituting the dimer, and $V_{nm}=\lambda$, the `tunnel-hopping amplitude' between the stub-atoms.  For the diamond-stub geometry we choose all the nearest neighbor hopping integrals as  $V_{nm}=L$. The presence of a magnetic flux trapped inside the diamond plaquette breaks the time-reversal symmetry locally, as the electron hops along an arm around the plaquette. As a result, we will associate a Peierls’ phase factor~\cite{hofs} $\theta = \frac{2\pi\Phi}{4 \Phi_{0}}$ with $L$, so that the `forward' and the `backward' hopping will eventually be  $ L e^{\pm i \theta}$. The nearest neighbor hopping is simply `L' when the hopping is taking place in a flux-free zone. 
 
 The amplitude of the wave function $\psi_n$ on any $n$th site obeys the time-independent Schr\"{o}dinger equation. $\psi_n$ satisfies the discrete version of the Schr\"{o}dinger equation, the so called `difference equation', viz,
\begin{equation} 
(E - \epsilon) \psi_n = \sum_m V_{nm} \psi_m
\label{diff}
\end{equation}
where, $E$ is the corresponding energy eigenvalue. In the following subsection, we will exploit Eq.~\eqref{diff} to understand the basic idea of generating extended eigenstates and a continuous spectrum.\\


\begin{figure}[ht]
\centering
(a)\includegraphics[width=.42\columnwidth]{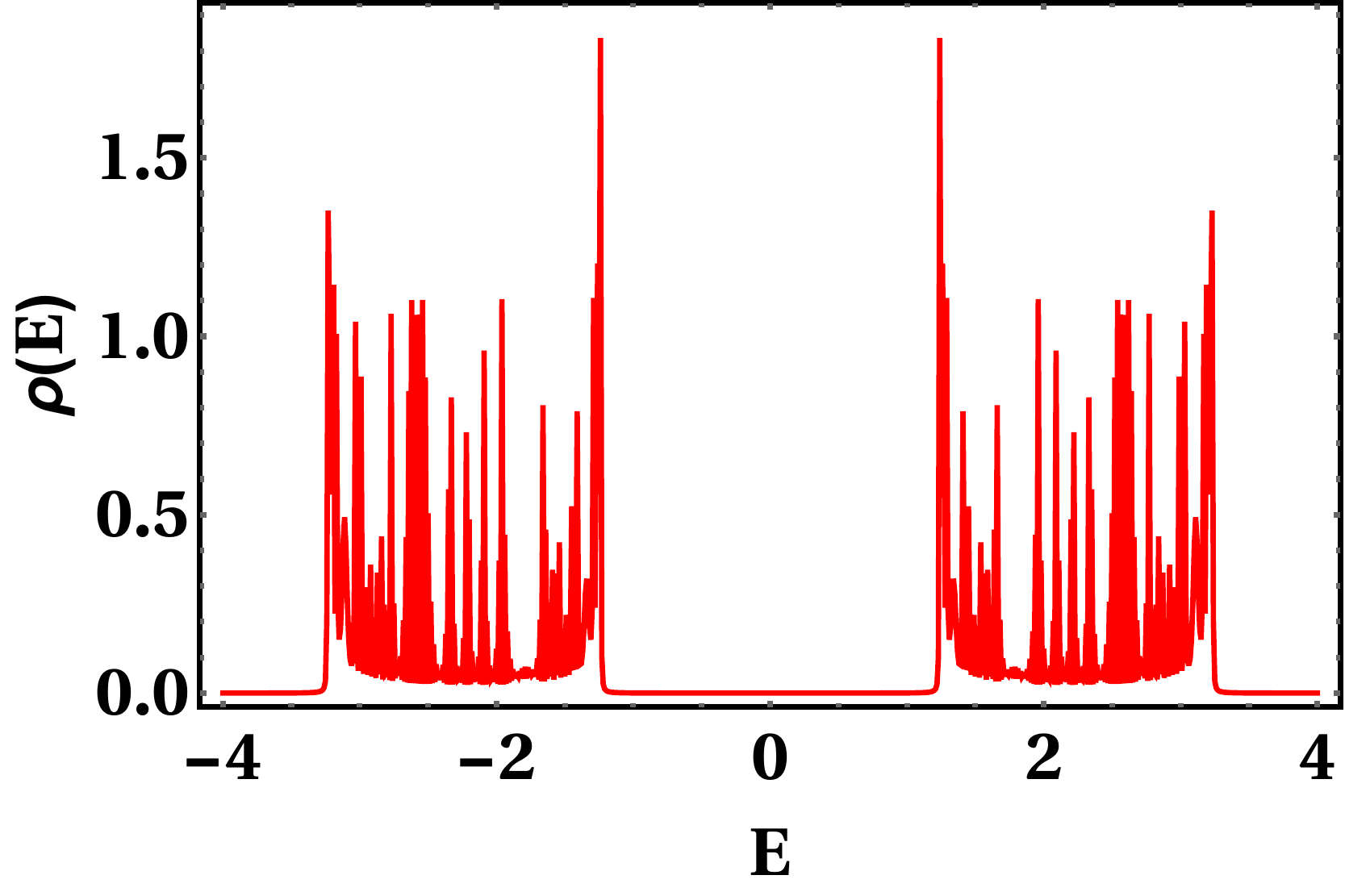}
(b)\includegraphics[width=.42\columnwidth]{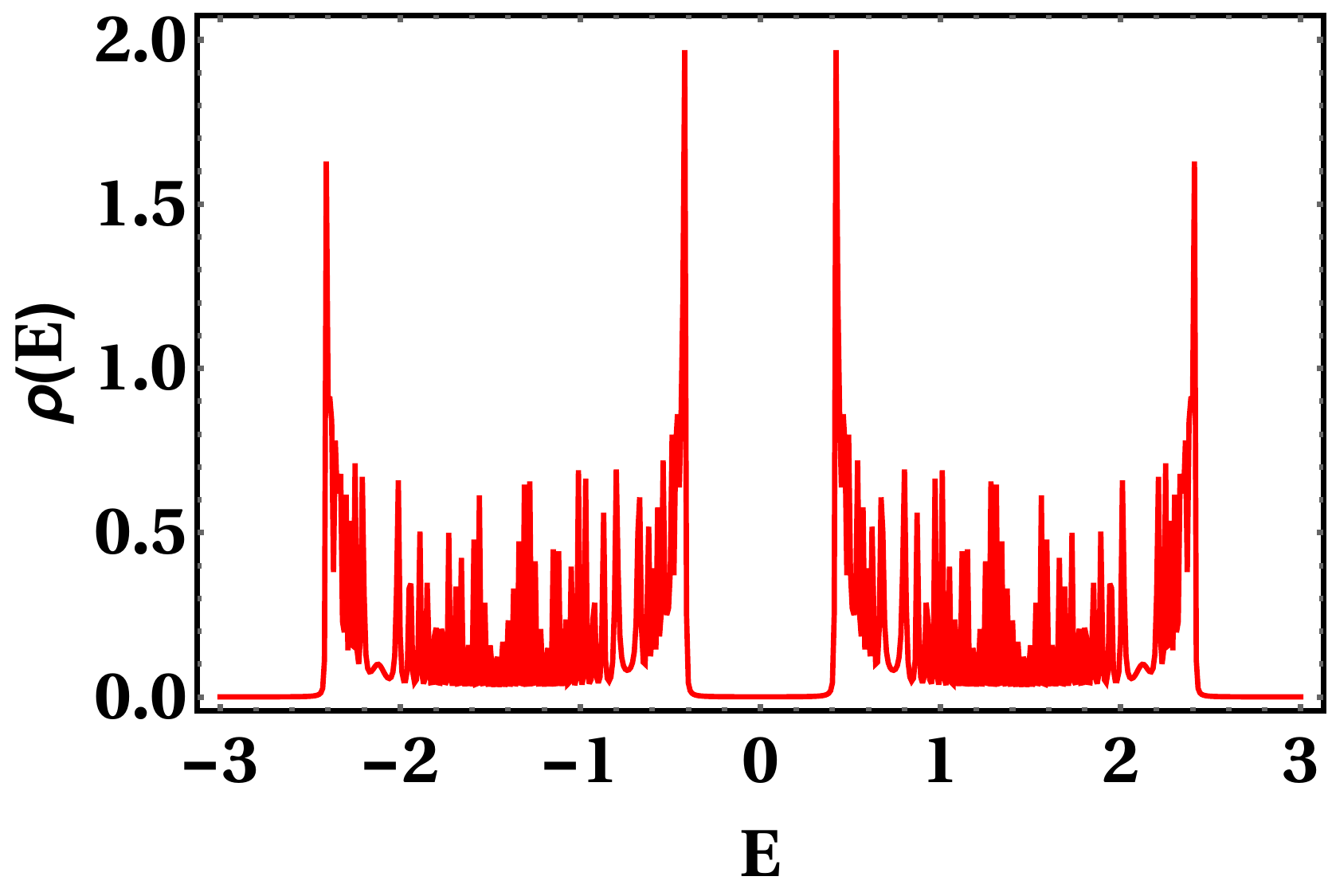}
(c)\includegraphics[width=.42\columnwidth]{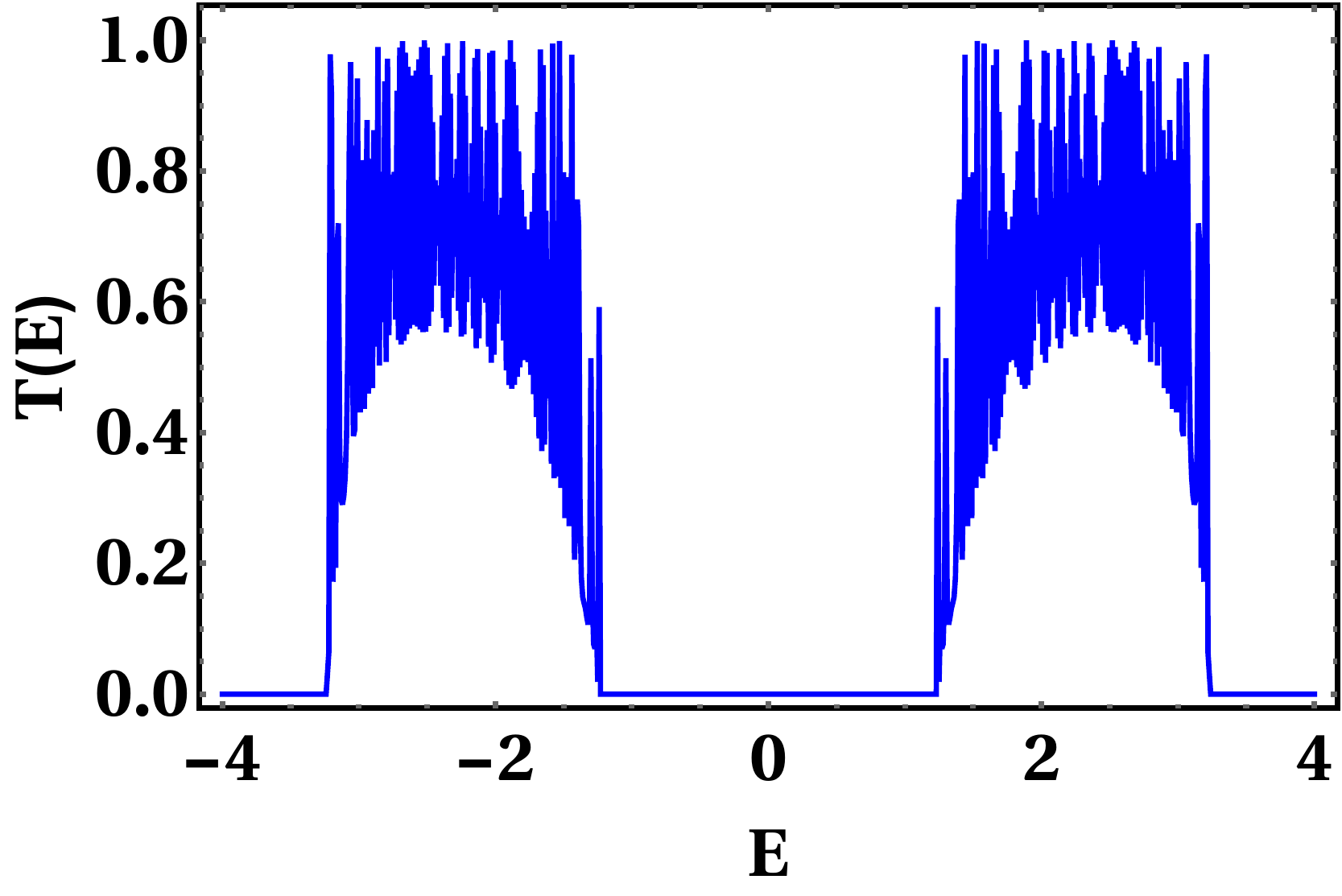}
(d)\includegraphics[width=.42\columnwidth]{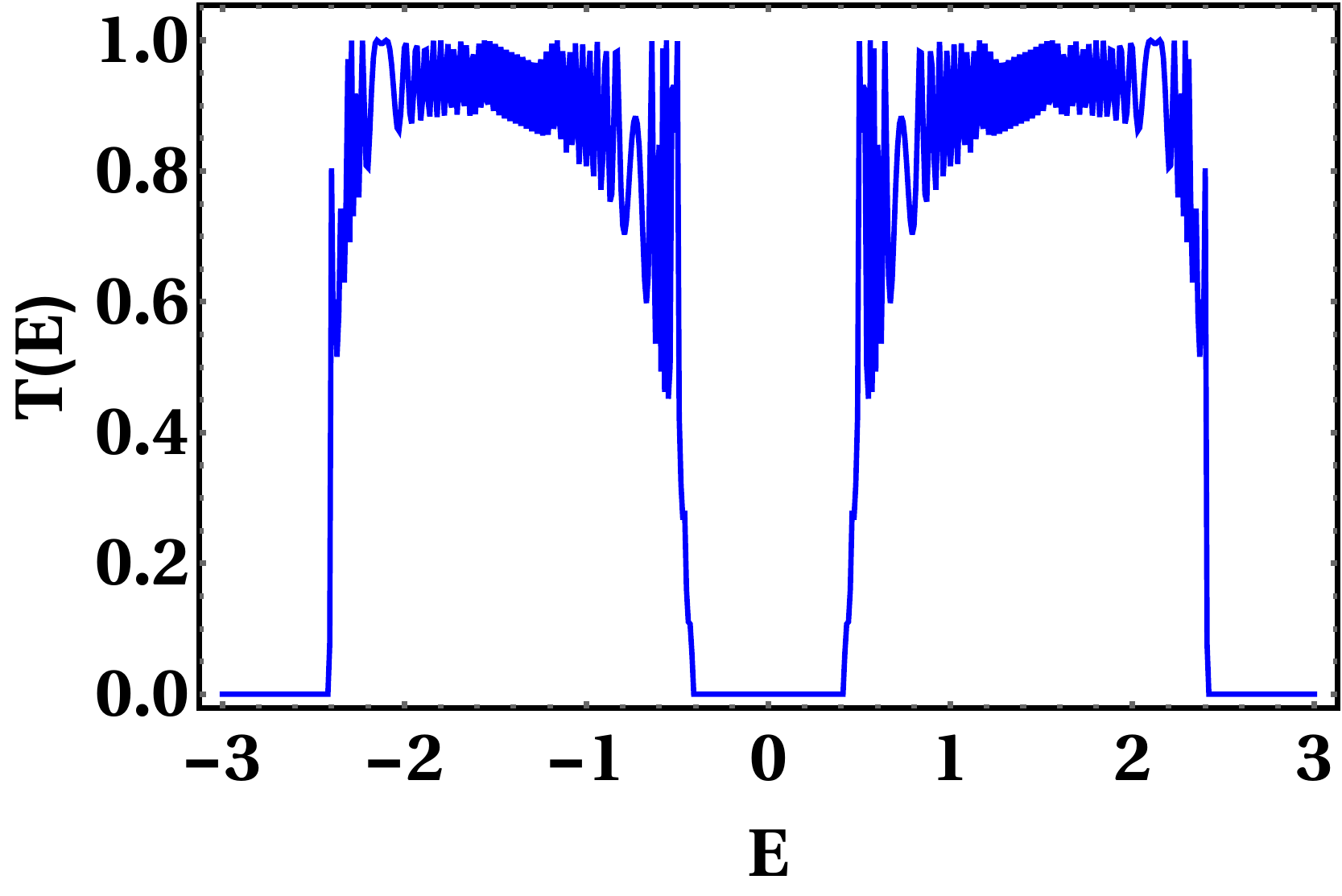}

\caption{(Color online) Variation of the average density of states (a,b) (outlined in Appendix A) and transmission co-efficient (c,d) (detailed in Appendix B) with energy for the Fibonacci array of `dimer-stub' ($290$ sites) (a,c), and `diamond-stub' ($468$ sites) (b,d) networks respectively. The parameters are chosen as, (a,c): $\epsilon = 0$, $L = 1$, $ S = \sqrt{5} $, $\lambda = 2$, and (b,d): $\epsilon = 0$, $L = 1$, $\Phi = \frac{1}{4}\Phi_0$. }  
\label{dos-trans}
\end{figure}

\subsection{Commuting transfer matrices: the extended eigenfunctions}

First, we decimate out the red colored atomic sites by expressing the amplitude of the wave function at the side coupled red colored site(s), in both Fig.~\ref{fig}(a) and (b), in terms of the amplitude at the base (blue) atom(s). The set of Eqs.~\eqref{diff} is used for this purpose. The resulting lattice is a purely one dimensional Fibonacci chain, depicted in Fig.~\ref{fig}(c). We use a single diagram to discuss both the dimer-stub and the diamond-stub examples. On this chain, we identify three kinds of sites, viz, $\alpha$ (green colored, flanked by an $LL$ pair), $\beta$ (yellow dots, flanked by an $LS$ pair) and $\gamma$ (orange dots, sitting between an $SL$ pair). Each of these sites $\alpha$, $\beta$ and $\gamma$ now have an energy-dependent on-site potential. The nearest neighbor hopping integrals in Fig.~\ref{fig}(c) are marked by $L$ and $S$. 

For the dimer-stub case (Fig~\ref{fig}(a)) the effective (renormalized) on-site potential of sites of type $\alpha$ is $\epsilon_{\alpha} = \epsilon+\frac{\lambda^2}{(E-\epsilon)}$, while $\epsilon_{\beta} = \epsilon_\gamma=\epsilon $, and remain unaffected by the decimation. The nearest neighbor hopping integrals $L$ and $S$ in this case are free from any energy dependence.

For the diamond-stub Fibonacci lattice, the `renormalized' on-site potentials are, $\epsilon_{\alpha} = \epsilon+\frac{L^2}{(E-\epsilon)}$, and $\epsilon_{\beta(\gamma)} =  \epsilon+\frac{2 L^2}{(E-\epsilon)}$ respectively. The nearest neighbor hopping amplitude $S$ for this case now has an energy dependence, and is given by, 
 $ S = \frac{ 2 L^2 \cos~(\frac{\pi \Phi}{\Phi_0})}{(E-\epsilon)}$. The other hopping amplitude $L$ remains energy independent.  

The decimation, in either case, allows us to deal with a purely one dimensional array of atoms with modified (renormalized) on-site potentials (for some of the sites in the dimer-stub situation, and for all the sites in the diamond-stub case), and corresponding modifications in the hopping integrals. The decimation method is already known to be useful in several other studies~\cite{tremblay,levstein}. However, we use it here just to remove a subset of the degrees of freedom, and use it once.

With the help of Eq.~\eqref{diff} it is now easy to relate the amplitudes of the wave function $\psi_N$ and $\psi_{N - 1}$ at sites $N$ and $N-1$ respectively to the amplitudes at the `initial' sites $\psi_1$ and $\psi_0$~\cite{kohmoto,lu} on this `modified' one dimensional chain through a matrix equation, 
\begin{equation}
\left[ \begin{array}{cc}
\psi_{N}\\
\psi_{N-1}\end{array}\right] = \prod_{n=N-1}^1\mathcal{M}_n 
 \left[\begin{array}{cc} \psi_{1}\\
\psi_{0}\end{array}\right]
\label{transfer-mat}
\end{equation}
Where $N$ is the total number of atomic sites and the transfer matrix $\mathcal{M}_n$ is given by,
\begin{equation}
    \mathcal{M}_n = \left[ \begin{array}{cccccccccccccccc}
 \frac{E-\epsilon_n}{V_{n,n+1}} & -\frac{V_{n,n-1}}{V_{n,n+1}}\\
 1 & 0\\ 
\end{array} \right ]
\end{equation}

In our case, the hopping amplitudes $V_{n,n \pm 1}$ are either $L$ or $S$ depending upon the bond connecting the neighboring sites on the effective one dimensional Fibonacci chain in Fig.~\ref{fig}(c) and $\epsilon_n = \epsilon_\alpha$, or $\epsilon_\beta$ or $\epsilon_\gamma$ depending on the site.

In the one dimensional chain in Fig.~\ref{fig}(c), whether it is obtained from a dot-stub geometry or a diamond-stub one, the transfer matrix corresponding to the $\alpha$ site, $\mathcal{M}_{\alpha}$ is written as,
\begin{eqnarray}
\mathcal{M}_{\alpha} & = & \left[ \begin{array}{cc}
 \frac{(E-\epsilon_{\alpha})}{L} & -1 \\
1 & 0 \\
\end{array}
\right ] \nonumber \\
\end{eqnarray}
The transfer matrix for `$\beta\gamma$' doublet is calculated by the product of the transfer matrices of $\gamma$ and $\beta$ sites, viz $\mathcal{M}_{\beta \gamma} =\mathcal{M}_{\gamma}\mathcal{M}_{\beta}$, where,

\begin{eqnarray}
\mathcal{M}_{\beta\gamma} & = & \left[ \begin{array}{cc}
 \frac{(E-\epsilon_{\gamma})(E-\epsilon_{\beta})}{L S} - \frac{S}{L} & -\frac{(E-\epsilon_{\gamma})}{S} \\
\frac{(E-\epsilon_{\beta})}{S} & -\frac{L}{S} \\
\end{array}
\right ] \nonumber \\
\end{eqnarray}

The commutator $\left[\mathcal{M}_{\alpha},\mathcal{M}_{\beta\gamma}\right]$ gives us the all important observation~\cite{biplab1,atanu}.                                                                              It turns out to be
\begin{equation}
\left[ \mathcal{M}_{\alpha}, \mathcal{M}_{\beta\gamma}\right]_{1(2)} = \left[ \begin{array}{cccccccccccccccc}
 0 & \Gamma_{1(2)}\\
\Gamma_{1(2)} & 0\\
\end{array}
\right ] 
\end{equation}

 The subscripts `$1$' and `$2$' designate the dimer-stub and the diamond-stub cases respectively.  $\Gamma_{1(2)}$ are the corresponding non-zero off-diagonal elements of the commutator.  They are given by,
 \begin{eqnarray}
     \Gamma_{1} &=&  \frac{\lambda^2 - (S^2-L^2)}{LS}\nonumber \\\
     \Gamma_{2} &=& -\frac{L \cos~(\frac{2 \pi \Phi}{\Phi_0}) \sec~(\frac{\pi \Phi}{\Phi_0})}{(E-\epsilon)}
     \label{commu}
 \end{eqnarray}
From the first of the set of Eqs.~\eqref{commu} it is clear that with an appropriate choice of hopping amplitude $\lambda = \sqrt{S^2-L^2}$ for `dimer-stub' Fibonacci sequence, $\Gamma_1=0$, and as a consequence, the commutator bracket becomes {\it zero} independent of the energy $E$ of the propagating particle. 

We just obtained the most important result for the dimer-stub lattice. This implies that, the clusters $\beta\gamma$ and the isolated $\alpha$ sites, no matter in which sequence (quasiperiodic or random) they were originally arranged can now be {\it re-arranged} at will. All such {\it new} arrangements will have an identical energy spectrum. So, the spectrum will be indistinguishable, for example, from the convolution of the spectra offered by an infinitely long {\it periodic} sequences of the $\beta\gamma$ dimers and an infinitely long periodic sequence of the $\alpha$ sites. Since both these cases separately represent chains with perfect translational order, the convolved spectrum must be of an absolutely continuous character with {\it all} the eigenstates extended. The end-to-end transmission coefficient for a finite, but long enough chain should be high, and close to unity (the scattering at the `lead-system' joining points however, prevents the transmission coefficient from reaching the exact value of unity). Usually, a Fibonacci lattice offers an insulating behavior with a very low transmission coefficient with a fractal character of the eigenstates. The {\it resonance} induced by a selective choice of the numerical values of the hopping amplitudes, as given in Eqs.~\eqref{commu}, should make the system metallic throughout the allowed ranges of energy eigenvalues.

Here, a comment regarding the long time dynamics is in order. For the dimer-stub model, the basic idea is that, at resonance, every individual cluster “$\beta\gamma$” or “$\alpha$”, can in principle, be arranged to form an infinitely long ‘periodic’ chain of their own. These systems, as individual chains are of course metallic in character, owing to their long range perfect translational order. Therefore the long time dynamics for each such periodic chain should be ballistic in nature. However, beyond the resonance condition of course, no such simplified remark can be made.


\begin{figure}[ht]
\centering
(a)\includegraphics[width=.8\columnwidth]{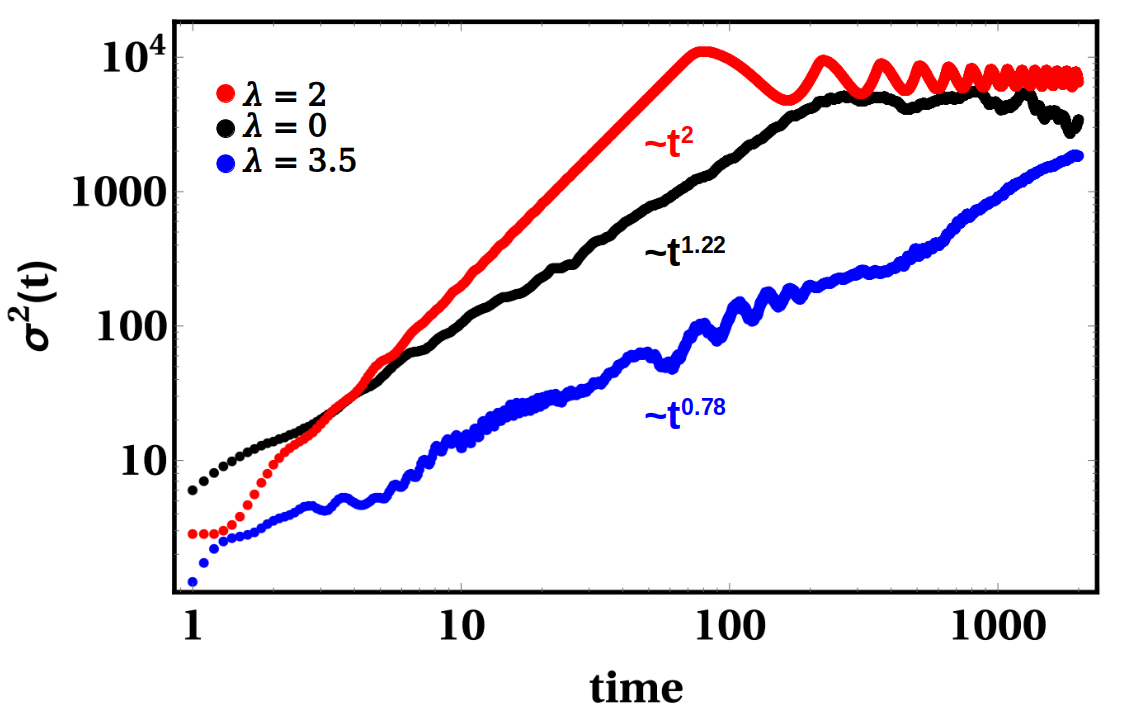}
(b)\includegraphics[width=.8\columnwidth]{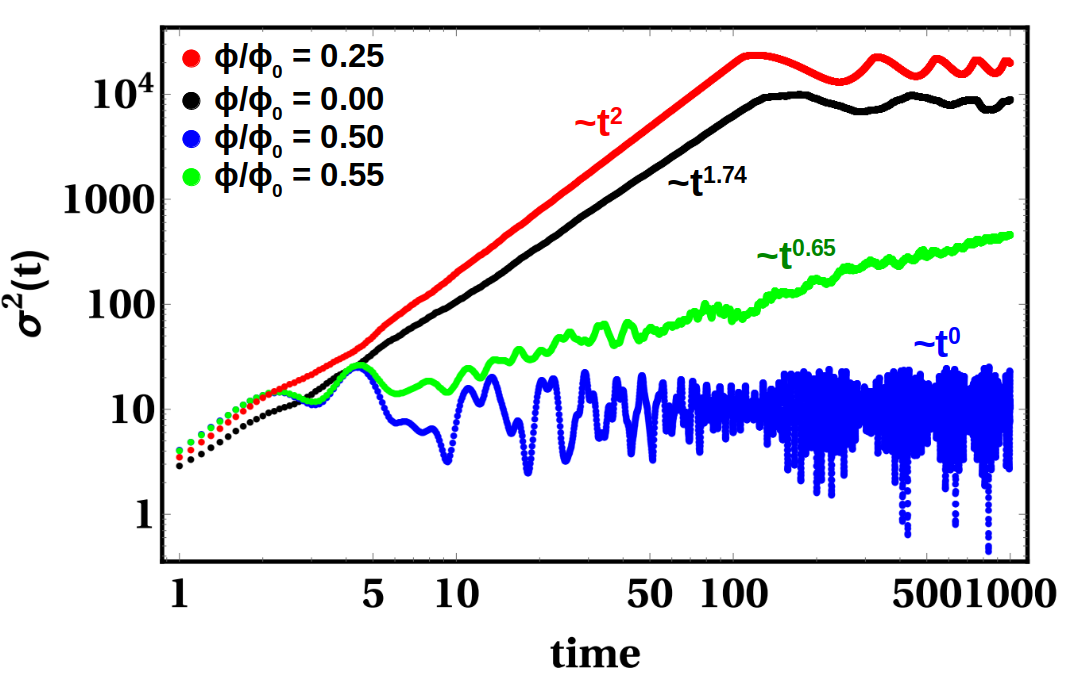}

\caption{(Color online) The mean-square displacement of a wave packet as a function of time (in a log-log scale) for the Fibonacci sequence of  (a) `dimer-stub' ($N = 290$ sites) with different values of the tunnel hopping amplitude $\lambda$, and (b) `diamond-stub' ($N = 468$ sites) with different values of trapped magnetic flux. The parameters are chosen as (a) $\epsilon = 0$, $L = 1$, $ S = \sqrt{5}$, $\lambda = 0 $ (black colored), $2$ (red colored), $3.5$ (blue colored) and for (b) $\epsilon = 0$, $L = 1$, $\frac{\Phi}{\Phi_0} = 0$ (black colored), $0.25$ (red colored), $0.50$ (blue colored), $0.55$ (green colored) respectively.}  
\label{msd1}
\end{figure}

In a similar line of arguments, we can appreciate that, for the diamond-stub Fibonacci array, reduced to a linear Fibonacci lattice of the kind of Fig.~\ref{fig}(c), a selection of magnetic flux $\Phi = \frac{1}{4}\Phi_0$ makes the commutator vanish, independent of energy $E$ again. As before, a swapping of the positions of the transfer matrices $\mathcal{M}_{\alpha}$ and  $\mathcal{M}_{\beta\gamma}$ is at our disposal. This implies we can again build up two infinitely long chains, one comprising the $\beta\gamma$ doublet, and the other one built out of the $\alpha$ sites alone. Each of these should give rise to a continuous energy spectrum populated by extended eigenfunctions only. Interestingly, this may be taken as an example of an insulator-to-metal transition driven by an external magnetic field only.

In Figs.~\ref{dos-trans} we present the average density of states (AVDOS) and the end-to-end transmission coefficient in the two cases discussed above. A summary of the procedure to evaluate the density of states is laid out in Appendix A for the sake of completeness. The scheme for calculating the transport is discussed in Appendix B. The AVDOS has been calculated by directly diagonalizing the Hamiltonian for a 290-sites dimer-stub chain and a 468-sites diamond-stub chain respectively. The gapless nature of the spectrum is already quite clearly visible (the wiggles are results of the finite system-size), and the corresponding transmission is practically perfect over the allowed energy bands. The results fully support our explanation.

We now proceed to study the quantum dynamics on these two systems. Now we have finite sized systems, and wish to see whether the observation about the extended character of the single particle eigenstates for the infinite systems is supported by the time evolution of a wave packet on such lattices, under the same resonance conditions.

\section{Quantum dynamics in dot-stub and diamond-stub Fibonacci sequences}

The wave packet of an electron at time $t = 0$ is written as, 
\begin{equation}
\ket{\Psi(0)} = \sum_{n} C_{n} (0) \ket{\psi_{n}(0)} 
\end{equation}
$C_n$ denotes the initial ($t=0$) probability amplitude of finding the particle found on the $n^{th}$ site. If the particle is released at some $m^{th}$ site, then $C_n (0) = \bra{\psi_{n}(0)}\ket{m}$. After some time, the time-evolved state is given by~\cite{katsanos, arka},
\begin{equation}
  \ket{\Psi(t)} = \sum_{n} C_n e^{-i E_{n} t} \ket{\psi_{n}(0)}
\end{equation}
We now study the fundamental quantities that govern the quantum dynamics of a wave packet on finite sized systems with the geometries shown in Fig.~\ref{fig}(a) and (b).

\subsection{Mean Square Displacement}

The mean square displacement (MSD) measures the spreading of the wave packet with time, which was initially at a particular site $m$ (say) at $t = 0$. It is defined as~\cite{katsanos,sougata1},
\begin{equation}
    \sigma^{2}(t) = \sum_{n} (n - m)^2 |\psi_{n}(t)|^2
    \label{msd}
\end{equation}


\begin{figure}[ht]
\centering
(a)\includegraphics[width=.44\columnwidth]{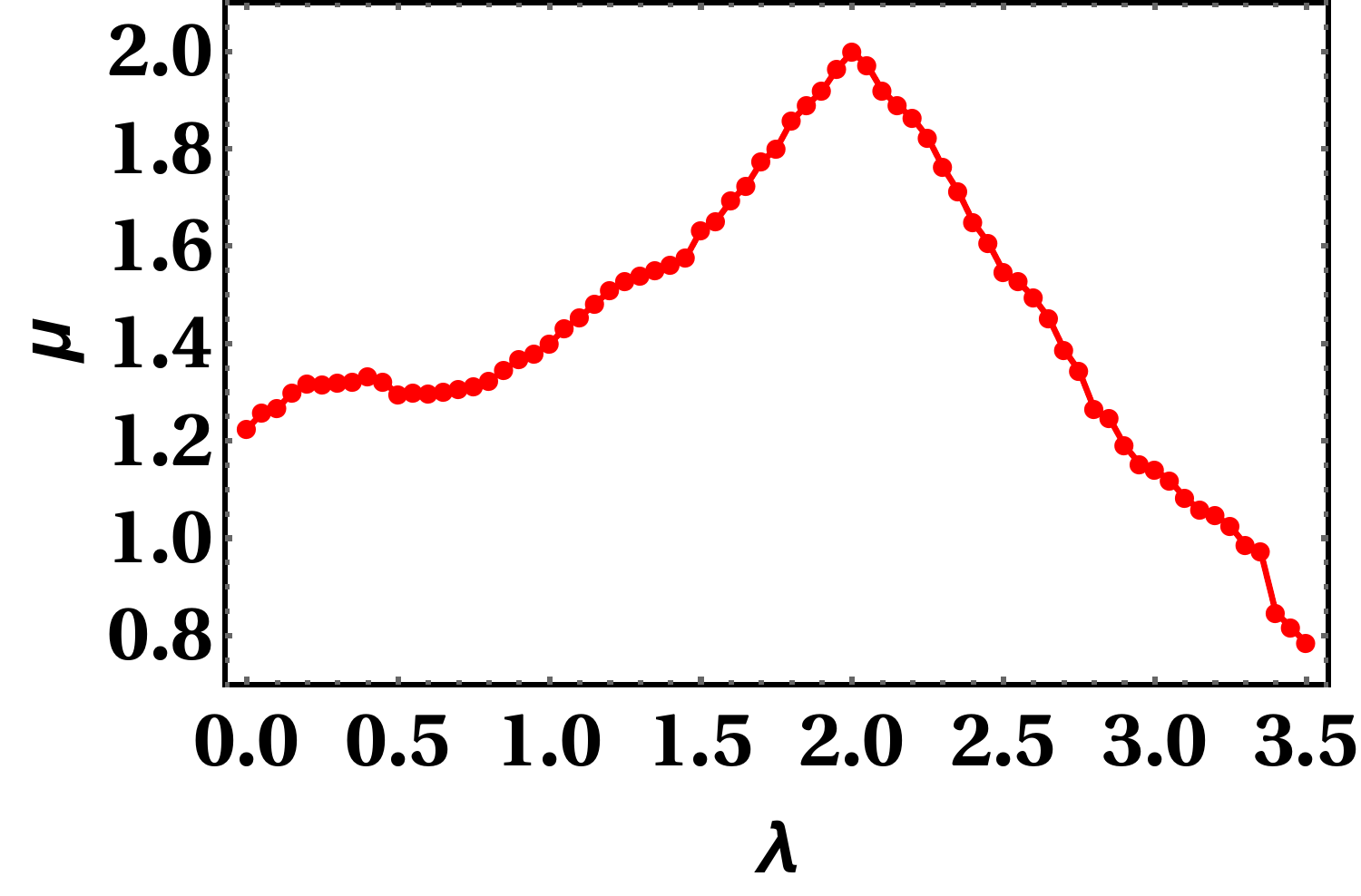}
(b)\includegraphics[width=.44\columnwidth]{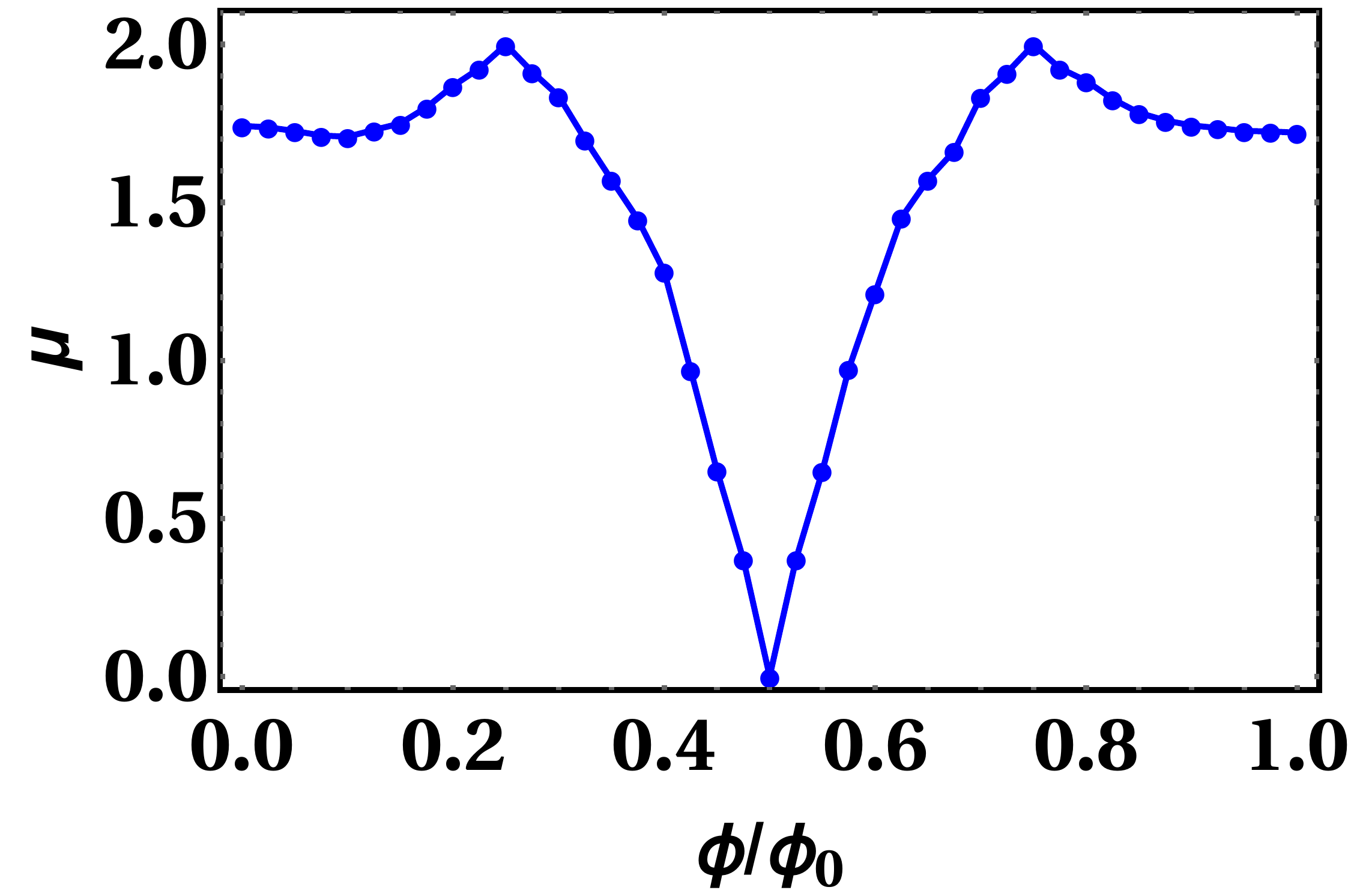}
\caption{(Color online) Variation of the scaling exponent $\mu$ (a) with the external hopping parameter $\lambda$ for the Fibonacci sequence of `dimer-stub' ($ 290$ sites), and (b) with the applied magnetic flux for `diamond-stub' ($468$ sites). The parameters are chosen as (a) $\epsilon = 0$, $L = 1$, $ S = \sqrt{5}$, and (b) $\epsilon = 0$, $L = 1$. }  
\label{exponent1}
\end{figure}

 The behaviour of $\sigma^{2}(t)$ displays an asymptotic dependence on time in the long time limit, and is described by a power law~\cite{katsanos} form $\sigma^{2}(t) \sim t^{\mu}$. Depending upon the numerical values of this exponent $\mu$, the dynamics of quantum wave packets can be classified as sub-diffusive for $0 < \mu < 1$,  localized, for $\mu=0$~\cite{siegle,vollmer,bodrova}, whereas $\mu = 1$, $2 > \mu > 1 $, and $\mu = 2$ indicate ordinary diffusion, super diffusion, and ballistic motion respectively~\cite{katsanos}.
 
 To see how the MSD varies in our chosen lattices when the resonance or off-resonance conditions are imposed, we worked with the original lattices shown in Fig~\ref{fig}(a) and (b) with the total number of sites $N =290$ and $N = 468$ respectively.
 At time $ t = 0$, the wave packet is released at some atomic site (say $m$). By calculating the MSD we can estimate its spreading in both resonance and off-resonance conditions if the system is allowed to evolve with time. At $t = 0$, we released the wave packet around the middle section, at  $m = 141$ and $m=234$ for the `dimer-stub' and the `diamond-stub' lattices respectively. The MSD was calculated as a function of time using  Eqs.~\eqref{msd}, and by directly diagonalizing the corresponding Hamiltonian matrix.
 
 The variation of MSD with time in a Fibonacci array of the `dimer-stub' quantum network is Shown in Fig.~\ref{msd1}(a). Here we have chosen the hopping integral along the horizontal bonds on the backbone as $L = 1$, and $S = \sqrt{5}$. The tunnel hopping amplitude between the stub-atom (red) and the base atom (blue), denoted by $\lambda$, is varied.  $\lambda = 2$ is our so called `resonance' condition. At the resonance condition, the infinite system was shown to exhibit two absolutely continuous subbands, and we examine what happens to the long time dynamics when we have a finite sized system.


\begin{figure}[ht]
\centering
(a)\includegraphics[width=.74\columnwidth]{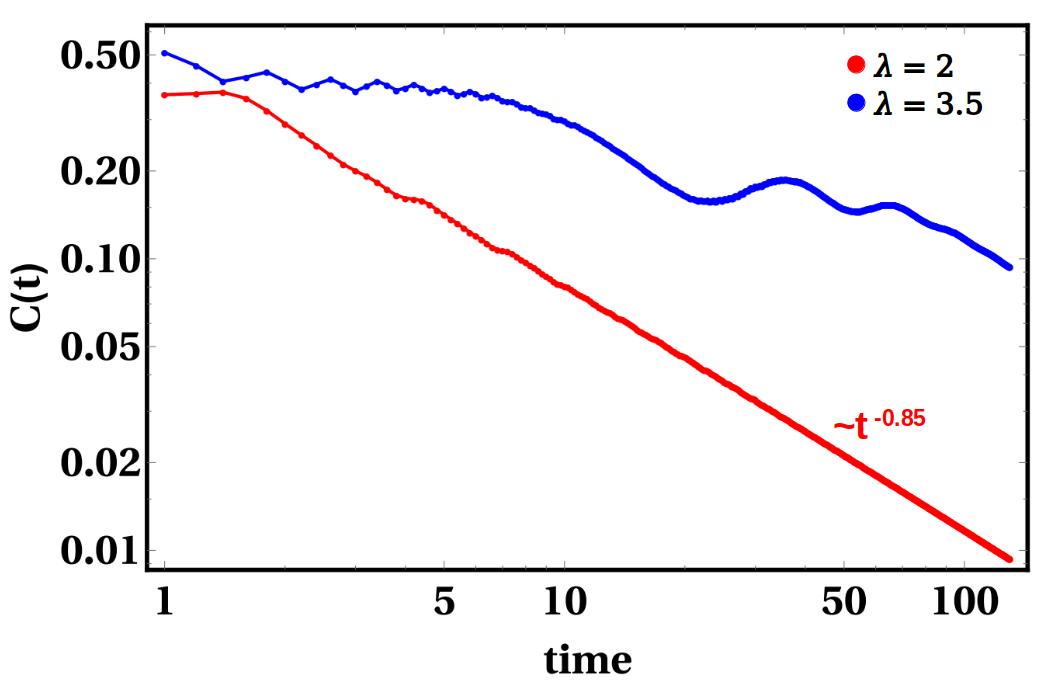}
(b)\includegraphics[width=.74\columnwidth]{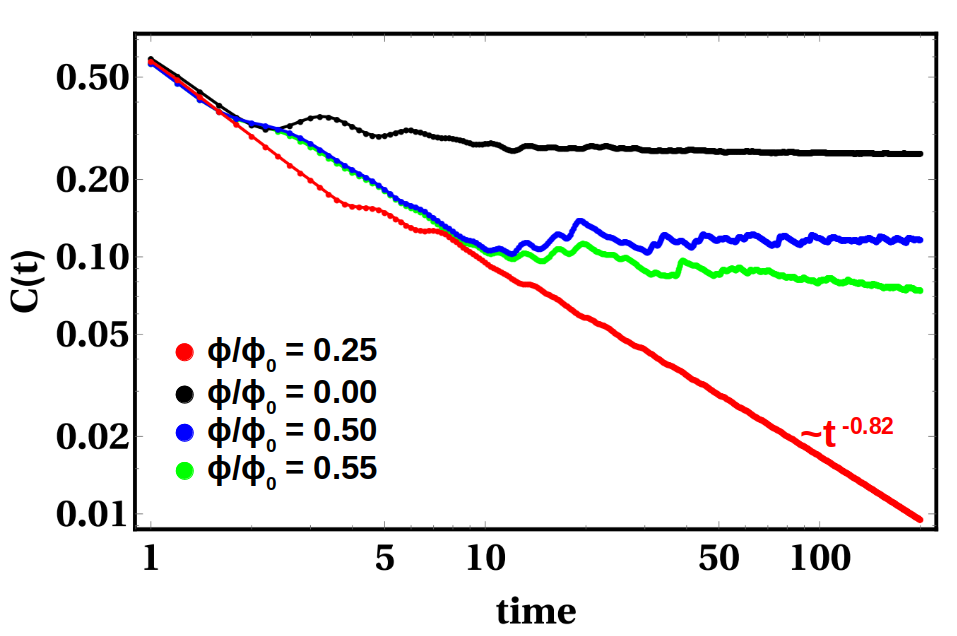}
\caption{(Color online) The temporal auto-correlation function is plotted against time using a log-log scale, for the 
Fibonacci sequence of (a) `dimer-stub' ($N = 290$ sites) with different values of the external hopping parameter $\lambda$ and (b) `diamond-stub' ($N = 468$ sites) with different values of applied magnetic flux. The parameters are chosen as (a) $\epsilon = 0$, $L = 1$, $ S = \sqrt{5}$, $\lambda =  2$ (red colored), $3.5$ (blue colored) and for (b) $\epsilon = 0$, $L = 1$, $\frac{\Phi}{\Phi_0} = 0$ (black colored), $0.25$ (red colored), $0.50$ (blue colored), $0.55$ (green colored)  respectively. }  
\label{taf1}
\end{figure}


\begin{figure*}[ht]
\centering

(a)\includegraphics[width=0.74\columnwidth]{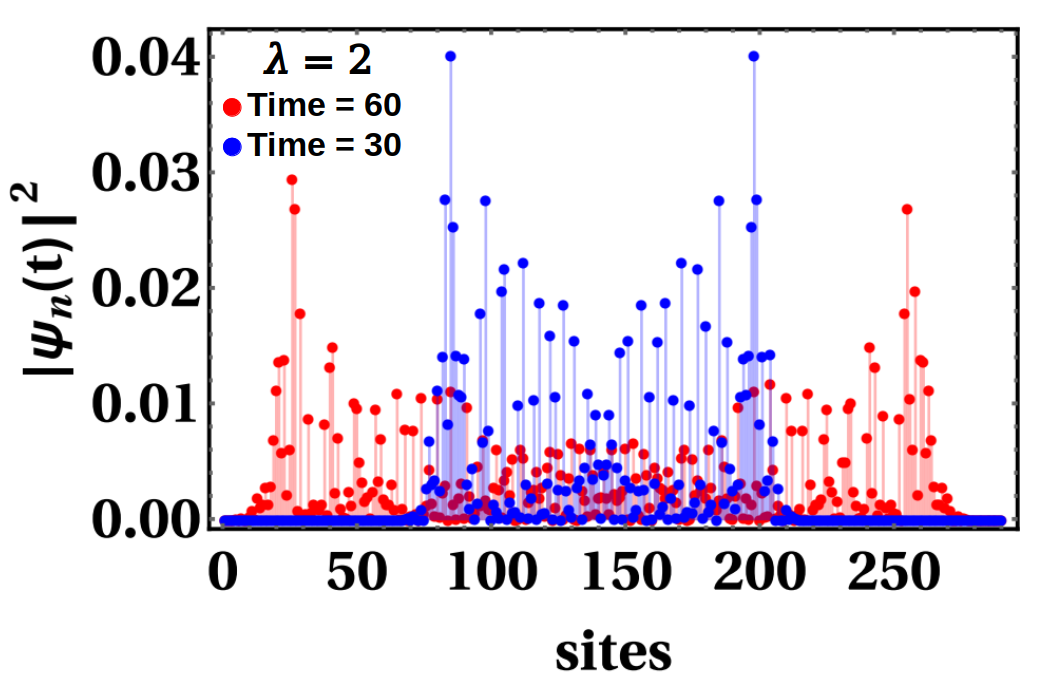}
(b)\includegraphics[width=0.74\columnwidth]{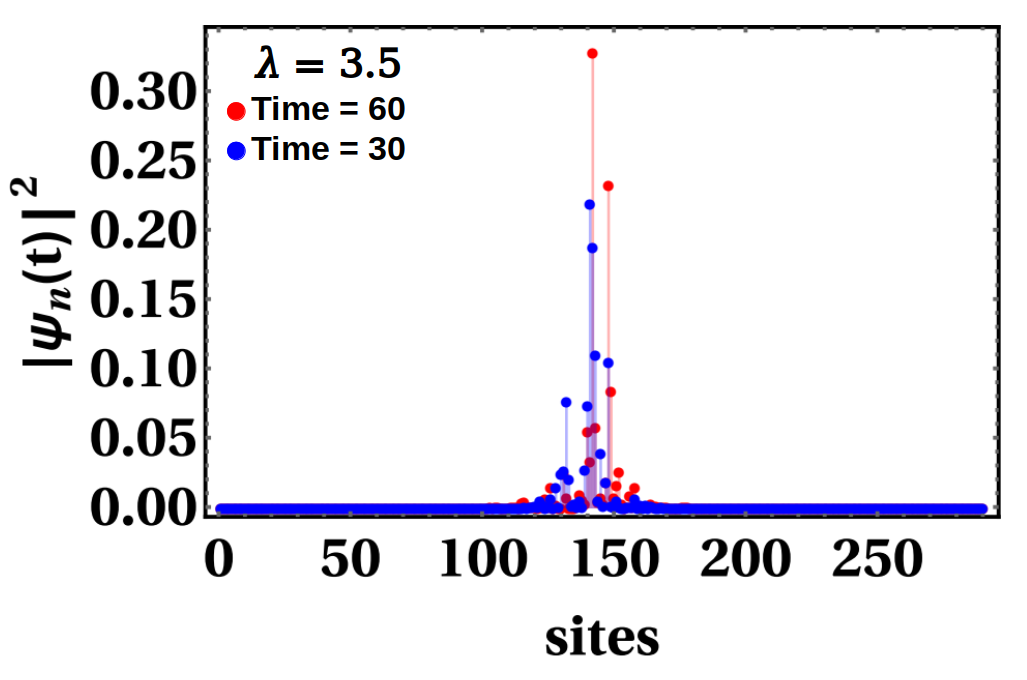}
(c)\includegraphics[width=0.74\columnwidth]{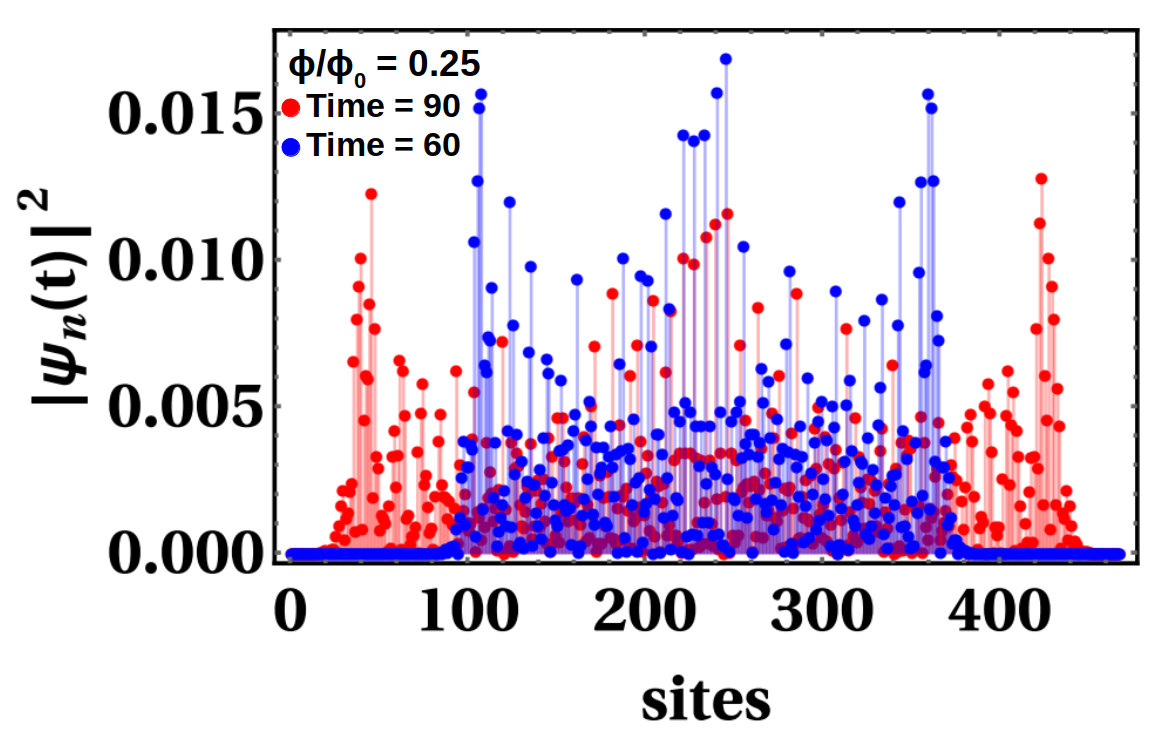}
(d)\includegraphics[width=0.74\columnwidth]{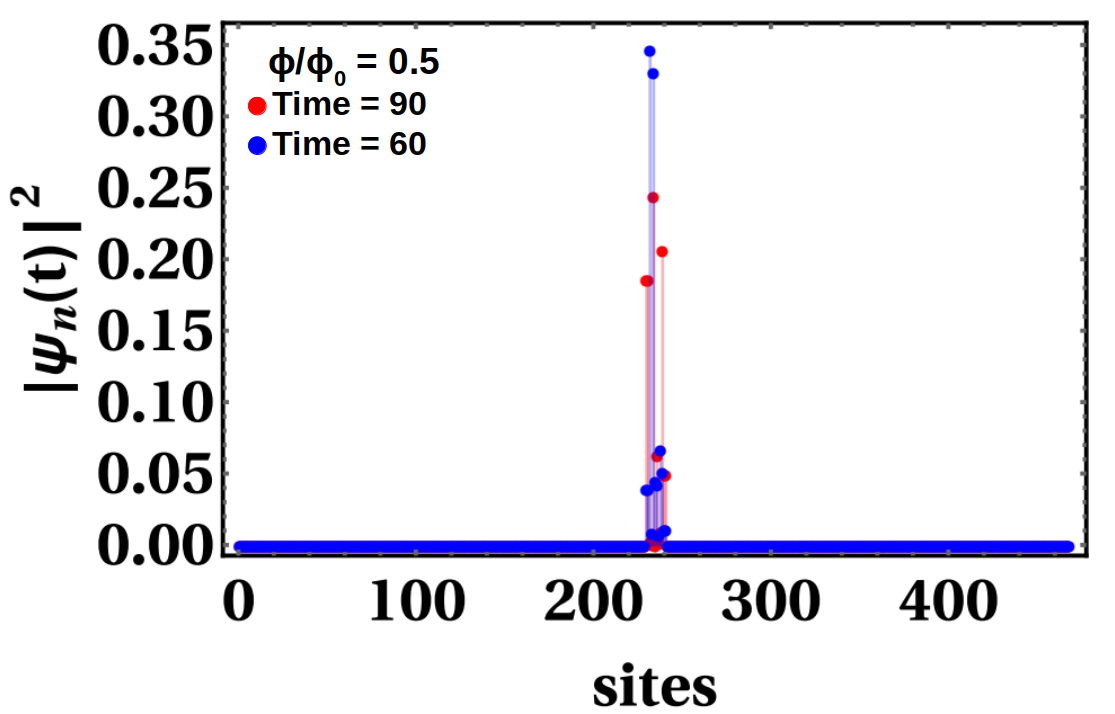}

\caption{(Color online) The variation of the spatial distribution of the wave packet in a Fibonacci sequence of (a,b) the dimer-stub lattice ($290$ sites), for time $t = 30$ (blue colored) and $60$ (red colored) with different values of the tunnel hopping (a) $\lambda = 2$, (b) $\lambda = 3.5$, and (c,d) for the diamond-stub lattice ($ 468$ sites) for time $t = 60$ (blue colored) and $90$ (red colored) with different values of applied magnetic flux (c) $\Phi = \frac{1}{4}\Phi_0$, (d) $\Phi = \frac{1}{2}\Phi_0$. Initially, we put the wave packet on the $141^{th}(234^{th})$ site for dimer-stub (diamond-stub) systems respectively. The hopping integrals are chosen as (a,b) $L = 1$, $ S = \sqrt{5}$, (c,d) $L = 1$, and all on-site potentials are set as zero. }  
\label{psi-n-1}
\end{figure*}
 
 When $\lambda = 0$ the system boils down to a linear Fibonacci array of $L$ and $S$ with isolated uncoupled dots (red colored) on one side. The long time beahvior of the wave packet shows a power law $ \sim t^{1.22}$, obtained using a best fit. The system behaves super diffusively. With $\lambda = 3.5$ the sub-diffusive behaviour sets in as the asymptotic (in time) behavior gives the best fit a power law $\sim t^{0.78}$. The long time behavior when the tunnel coupling $\lambda=2$ is obtained using a best fit again, and the temporal behavior is found to be governed by a scaling law of $\sigma^2(t) \sim t^2$. 
 The appearance of a saturation (the red curve) simply indicates that beyond a certain time, the wave packet has spread over the entire lattice, and no further change in the MSD is expected. We have checked that with a larger system, the saturation is delayed and starts at a later time. The $t^2$ scaling of the MSD persists before the wiggling begins.
 This is clearly indicative of a ballistic motion, and we see that the characteristics of an infinite system is already reflected for a moderately large system size. 

In Fig.~\ref{exponent1}(a) the behaviour of scaling exponent $\mu$ is shown against the variation of $\lambda$. From the graph, it is clear that the wave packet dynamics can be tuned in the ballistic regime, diffusion or super diffusive regimes, or into a phase of sub-diffusion by controlling the stub-to-backbone coupling $\lambda$ appropriately.

The MSD as a function of time for the `diamond-stub' Fibonacci sequence is shown in Fig.~\ref{msd1}(b). Now the magnetic flux trapped in each diamond-cavity, plays the all important role to control the wave packet dynamics. Without any flux the MSD with time shows, in the long time limit, a power law $\sim t^{1.74}$. The system is found to be in a 
 super diffusive state. For an applied magnetic flux $\Phi = \frac{1}{4} \Phi_0$ in each diamond plaquette, the system achieves the resonance condition, and the scaling exponent  $\mu $ is $\approx 2$, corresponding to the ballistic motion. This again corroborates our analytical result using the commuting transfer matrix concepts, as outlined before. The time exponent $\mu$ is $\approx 0$ when each diamond cavity is trapped by a magnetic flux $\Phi = \frac{1}{2}\Phi_0$. This is essentially an example of an {\it extreme localization}~\cite{vidal}. In this case, the amplitudes of the single particle wave functions survive only at certain special points on the lattice say, at the top and bottom sites of the diamond, and the clusters of non-zero amplitudes are separated from another such set by sites where the wave amplitude vanishes. Deviating to a value close to $\Phi = \frac{1}{2}\Phi_0$, say $\Phi = 0.55 \Phi_0$ the power law becomes $\sim t^{0.65}$ which implies that the system is in the sub-diffusive regime.

The variation of scaling exponent $\mu$ against applied magnetic flux is plotted in  Fig.~\ref{exponent1}(b). The magnetic flux is tuned from $0$ to $1$ with an interval of $0.025$. At zero flux the system exhibits superdiffusive motion then the system shows its ballistics character when applied flux is  $\Phi = \frac{1}{4} \Phi_0$. Further increment of magnetic flux puts the system in sub-diffusive region via super(or ordinary) diffusive motion and a pinned localized state appears at $\Phi = \frac{1}{2} \Phi_0$.  Further increase in $\Phi$ (beyond $\Phi_0/2)$ we again observe an ordinary (super) diffusive behaviour, and the ballistics motion sets in at $\Phi = \frac{3}{4} \Phi_0$. To our mind, this is an interesting example of {\it flux drivem} re-entrant diffusive-ballistic-diffusive transition in quantum dynamics.


\begin{figure*}[ht]
\centering
(a)\includegraphics[width=0.62\columnwidth]{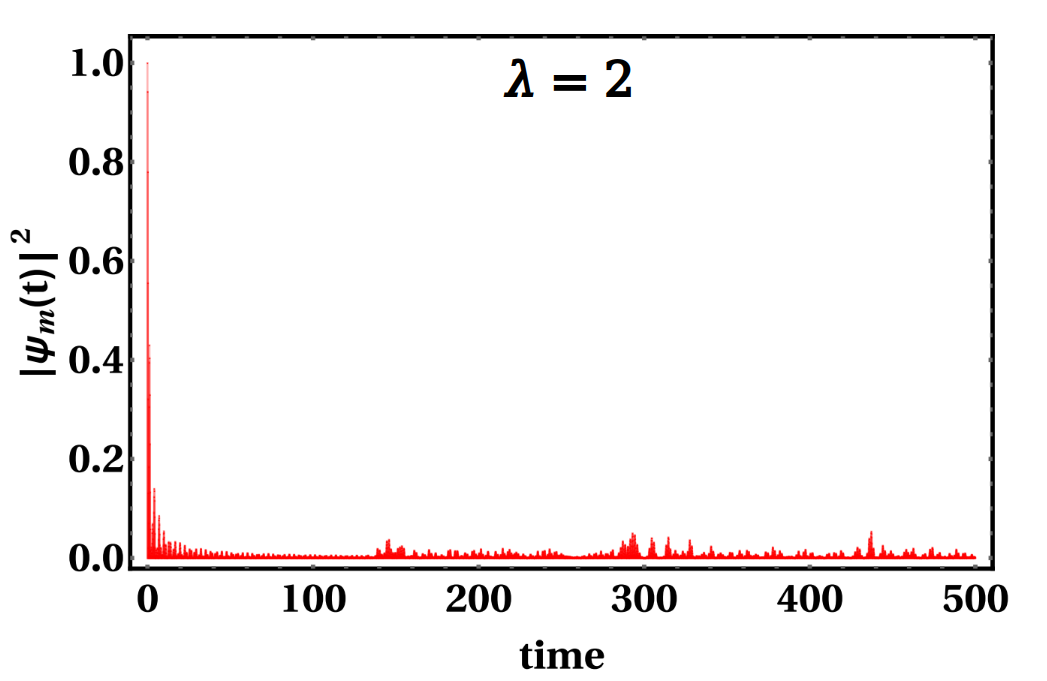}
(b)\includegraphics[width=0.62\columnwidth]{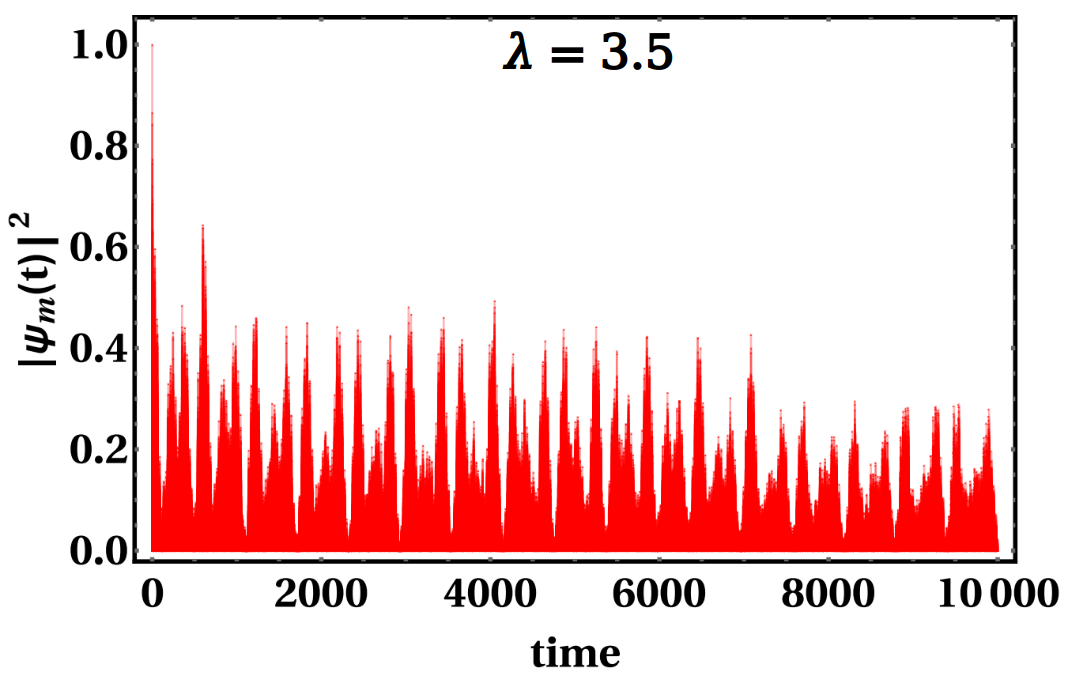}
(c)\includegraphics[width=0.62\columnwidth]{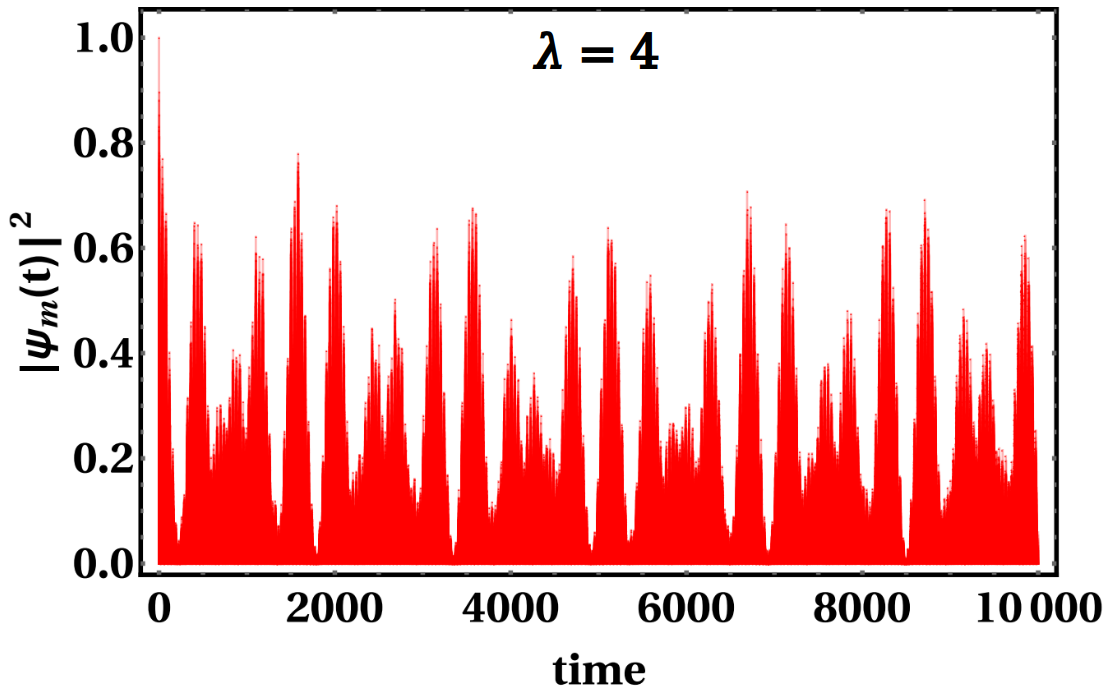}\\
(d)\includegraphics[width=0.62\columnwidth]{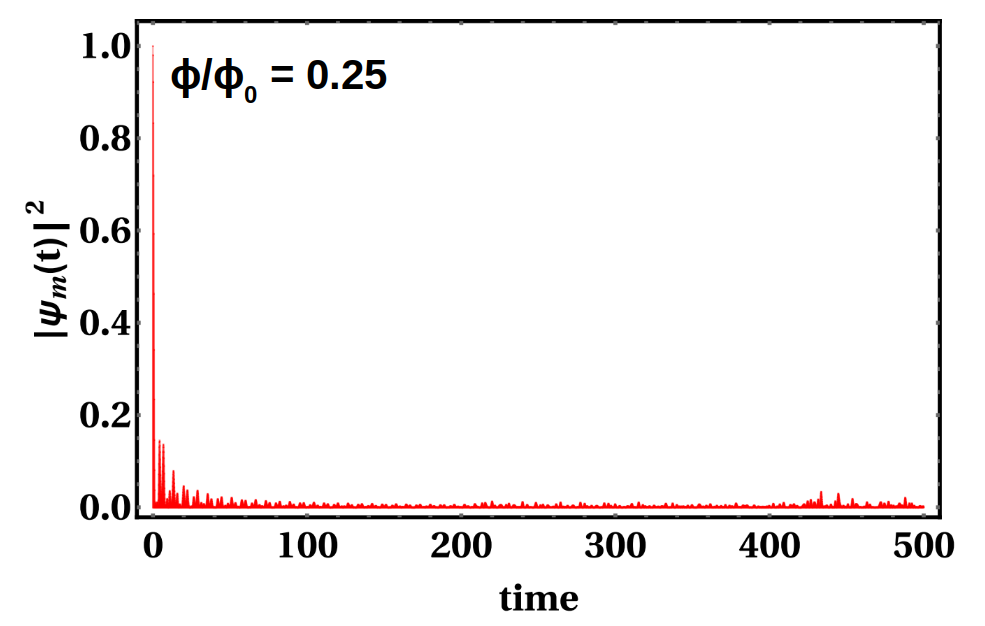}
(e)\includegraphics[width=0.62\columnwidth]{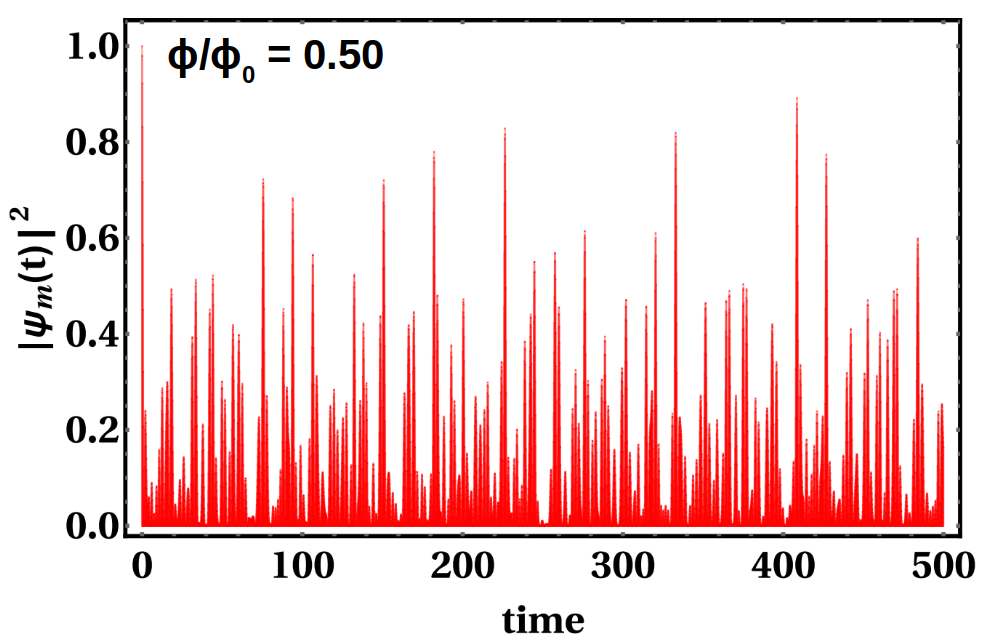}

\caption{(Color online)  The probability of finding the wavepacket at the initial site as a function of time for a Fibonacci sequence of (a,b,c)`dimer-stub' quantum network ($290$ sites) with different values of external parameter (a) $\lambda = 2$, (b) $\lambda = 3.5$, (c) $\lambda = 4$ and (d,e) for `diamond-stub' network ($ 468$ sites) with different values of applied magnetic flux (d) $\Phi = \frac{1}{4}\Phi_0$, (e) $\Phi = \frac{1}{2}\Phi_0$. Initially, we put the wave packet at the $141^{th}$ and at the $234^{th}$ site for `dimer-stub' and the `diamond-stub systems respectively. The hopping integrals are chosen as (a,b,c) $L = 1$, $ S = \sqrt{5}$, (d,e) $L = 1$, and all on-site potentials are set equal to zero. }
\label{return1}
\end{figure*}

\subsection{ The return probability and the temporal autocorrelation function }
 If a wave packet is initially released at a particular site $m$ at time $t = 0$, then at a later time $t$ the probability of finding the wave packet at site $m$ is given by $P_{m}(t) = |\Psi_{m}(t)|^2$. This is the return probability (RP). After evolving for a long time if RP at the initial site $m$ (and/or its neighborhood) is still finite, we get an indication of localization in general. For a delocalized phase, as the time flows on, the RP decays to zero~\cite{guan}. 

The temporal autocorrelation function (TAF) is a time-averaged representation of the return probability~\cite{katsanos,arka}.
\begin{equation}
    C(t) = \frac{1}{t}  \int_{0}^{t} |\Psi_{m}(t)|^2 \,dt 
    \label{eq-taf}
\end{equation}\\
The behaviour of TAF for large times is dictated by the power-law~\cite{katsanos} $C(t) = t^{-\delta}$. For a completely periodic system, this exponent $\delta$ is $\approx 1$ in the long time limit. 

Using Eq.~\eqref{eq-taf} we calculate $C(t)$ as a function of time for both the Fibonacci sequence of the `dimer-stub' and `diamond-stub' quantum networks. In Fig.~\ref{taf1} (a) the TAF is plotted, in a log-log scale, against time for the `dimer-stub' network and we find that, as we set $\lambda=2$ (the resonance condition) the decay of $C(t)$ after a long time exhibits a power-law-like form $t^{-\delta}$ with the exponent $\delta$ taking up a value close to $0.85$ when time is not so large. This is shown by the red line in Fig.~\ref{taf1} (a). Going to an even longer time, we have checked that the exponent comes even closer to unity. For $\lambda = 3.5$, a strong deviation from our resonance condition, the TAF has a very small decay-rate with time (blue colored line) which can be attributed again to the sub-diffusive character of the wave packet.

The variation of $C(t)$ with time for the Fibonacci sequence of `diamond-stub' is shown in Fig.~\ref{taf1}(b) when a constant magnetic flux is trapped in each diamond cavity. For $\Phi = \frac{1}{4}\Phi_{0}$ (resonance condition) after a large time it has a tendency to follow $C(t) \approx t^{-1}$ (red colored) whereas its decay for other values of the magnetic flux are depicted in the same figure shows the localized nature of the wave packet, a typical characteristic of the underlying quasiperiodic structure.

To substantiate the observations above, we measure the spatial variation of $ |\Psi_{n}(t)|^2 $ at different instants of time. In Fig.~\ref{psi-n-1} (a,b) the spatial distribution of a wave packet initially at $m = 141^{th}$ site at two different times $ t = 30$ (blue colored) and $60$ (red colored) is shown for the  `dimer-stub' Fibonacci sequence. At $\lambda = 2$ the wave packet spreads from its initial location throughout the chain as time progresses (Fig.~\ref{psi-n-1}(a)). This is the signature of a perfect de-localization.  As expected, there is little spreading of the wave packet with time when $\lambda = 3.5$ (Fig.~\ref{psi-n-1}(b)) owing to the sub-diffusive character of the state.

A similar plot for the `diamond-stub' network is shown in Fig.~\ref{psi-n-1}(c,d) for different values of magnetic flux. In Fig.~\ref{psi-n-1}(c) the trapped magnetic flux is $\Phi = \frac{1}{4}\Phi_0$, such that the system is in the resonance condition. The wave packet is released at $m=234$ at $t=0$. It is found to spread out over the entire system with time. Two such spread out distributions are illustrated site at time $t = 60$ (blue colored), and at  $t=90$ (red colored), whereas it is seen to be localized practically at its initial position even after a large time when applied flux is set at $\Phi = \frac{1}{2}\Phi_0$ (Fig.~\ref{psi-n-1}(d)) signifying an extreme localization, as discussed before.

We conclude this subsection by inspecting the RP itself as a function of time both for the Fibonacci sequence of `dimer-stub' and `diamond-stub' units. In Fig.~\ref{return1} (a,b,c) the return probability is plotted against time, for the `dimer-stub' network in resonance and off-resonance conditions. At resonance,  $\lambda = 2$, after a finite time there is no return probability. That is, the wave pack never comes back to its origin (Fig.~\ref{return1}(a)). Further increment of $\lambda$ ($\lambda = 3.5, 4$) makes the wave-packet spend more time along the `stub'  (as $\lambda$ is quite large now compared to $L$, that is, the hopping along the backbone). For $\lambda=3.5$, the wavepacket is in the sub-diffusive regime. Therefore, for long enough time, the profile of RP will eventually show a decay, implying an `escape' (though slowly) of the particle. As $\lambda$ becomes equal to $4$, the sub-diffusive regime still persists, but now the trend of the decay in RP is not as prominent as in the earlier case. The trends are clearly seen in Fig.~\ref{return1}(b) and (c).

The RP in the Fibonacci sequence of the `diamond-stub' network for different values of magnetic flux is shown in Fig.~\ref{return1} (d,e). Again, as expected, for $\Phi = \frac{1}{4} \Phi_0$ the `diamond-stub' geometry becomes equivalent to a perfectly periodic lattice, and the wavepacket spreads all over the lattice with time. A return to the origin with time is totally absent (Fig.~\ref{return1}(d)). A pinned localized character of the wave packet at $\Phi=\frac{1}{2}\Phi_0$ is evident, and is corroborated by the time evolution as shown in Fig.~\ref{return1}(e).



\begin{figure}[ht]
\centering
(a)\includegraphics[width=.8\columnwidth]{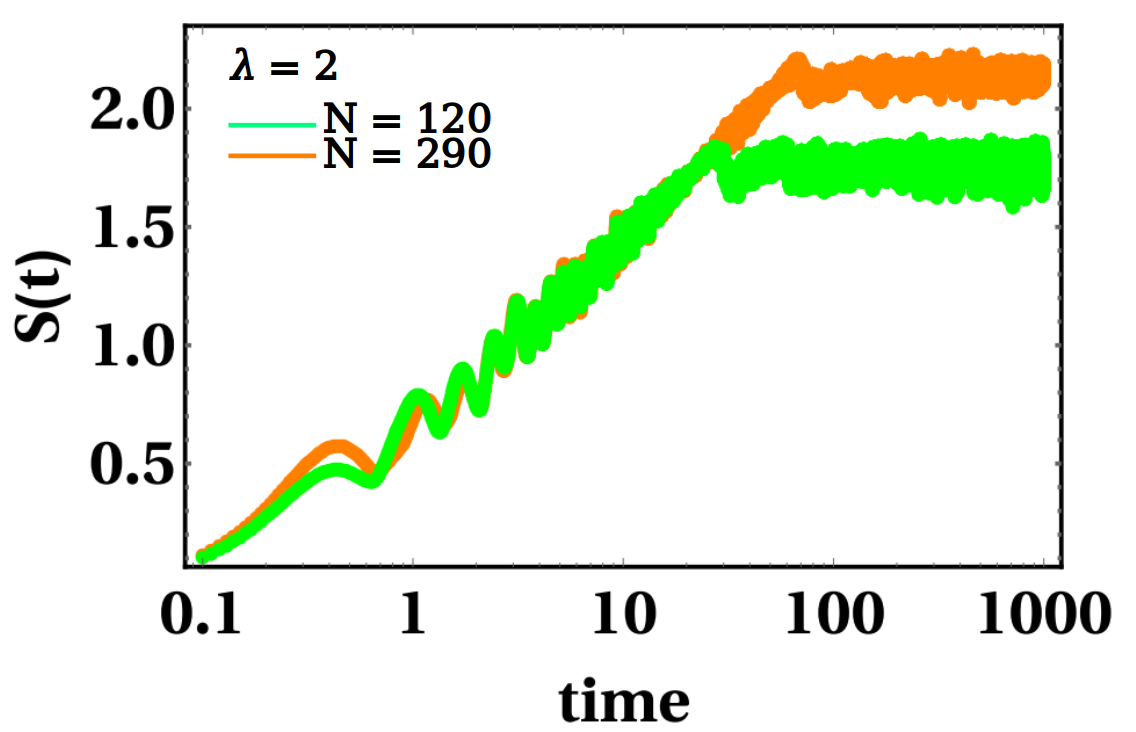}
(b)\includegraphics[width=.8\columnwidth]{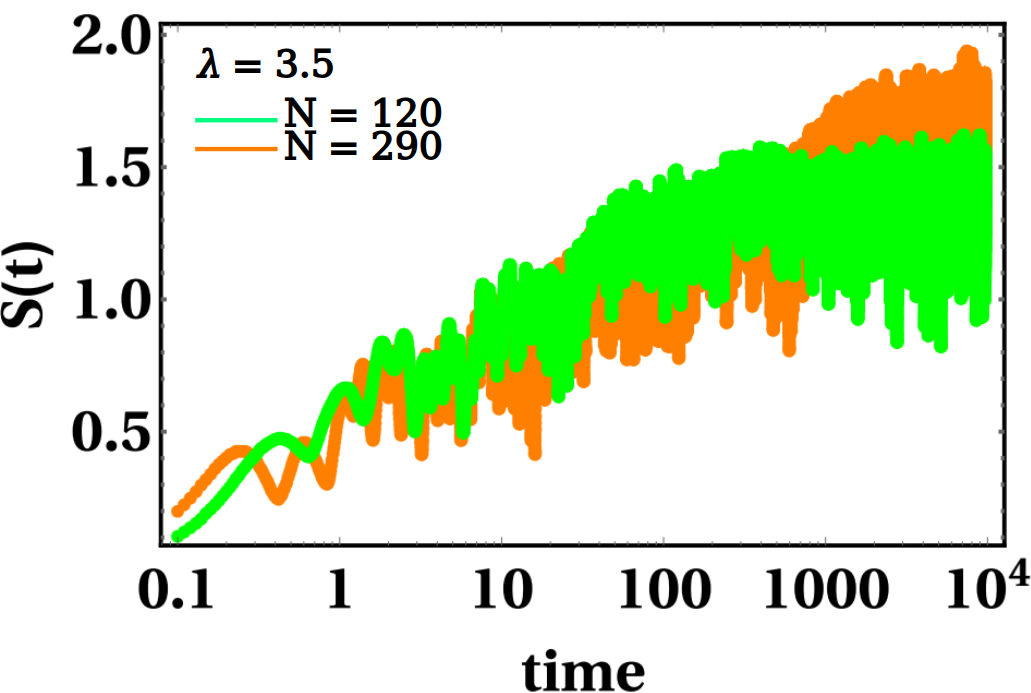}
\caption{(Color online)  The information entropy is plotted against time (in a log scale) for the Fibonacci sequence of the (a,b) `dimer-stub' quantum network with different values system size $N =290$ (orange colored) and $N = 120$ (green colored) when the external hopping parameter (a) $\lambda =2$, and (b) $\lambda = 3.5$ respectively. The hopping integrals are chosen as $L = 1$, $ S = \sqrt{5}$, and all on-site potentials are set as zero. The initial site where the wavepacket is released is chosen as $m = 141^{th}$ and $ m = 60^{th}$ for system size $N = 290$  and $N = 120$ respectively.}  
\label{entropy1}
\end{figure}

\subsection{Information Entropy}
Information entropy (IE) is a very popular measurement to detect (de)localization~\cite{ravi}. This is  written in the following form~\cite{katsanos, farzadian, coppola},
\begin{equation}
    S(t) = - \sum_{n} P_{n} \log P_{n}
\end{equation}
Where $P_n = |\Psi_{n}(t)|^2$ is the probability of finding the wave packet in its $n^{th}$ state with $0 \leqslant P_{n} \leqslant 1$ and $\sum_{n} P_{n} = 1$. As discussed in the previous subsection (Fig.~\ref{psi-n-1}(a), Fig.~\ref{psi-n-1}(c)) for a delocalized phase, the wave packet spreads throughout the lattice from its initial position $m$ as time evolves. As a result there is a finite probability of finding the wave packet on each site of the system. In the case of localization, the wave packet is still confined around  its initial location $m$ after a long time interval so that, very close to $m$ the probability is high, and in all other positions the probability is almost zero (Fig.~\ref{psi-n-1}(b), Fig.~\ref{psi-n-1}(d)).  Due to this type of probability distribution $S(t)$ increases monotonically with time ~\cite{katsanos} when the system is delocalized, and a saturation appears when the wavepacket fully spreads all over the system.  The value of the saturation-point increases linearly with the system size ~\cite {coppola, nakagawa} because, with an increasingly large system size, the wavepacket takes greater time to spread out. Hence, $S(t)$ shows a greater saturation value, and the saturation sets in at a later time.  On the other hand, the IE shows a very short time growth followed by saturation in a localized phase, and this saturation is not affected by the system size ~\cite {ravi,bardarson,coppola}. Because, for the localized phase the wavepacket is always centered around its initial location (no spreading), no matter for how long the system evolves with time. 


\begin{figure}[ht]
\centering
(a)\includegraphics[width=.8\columnwidth]{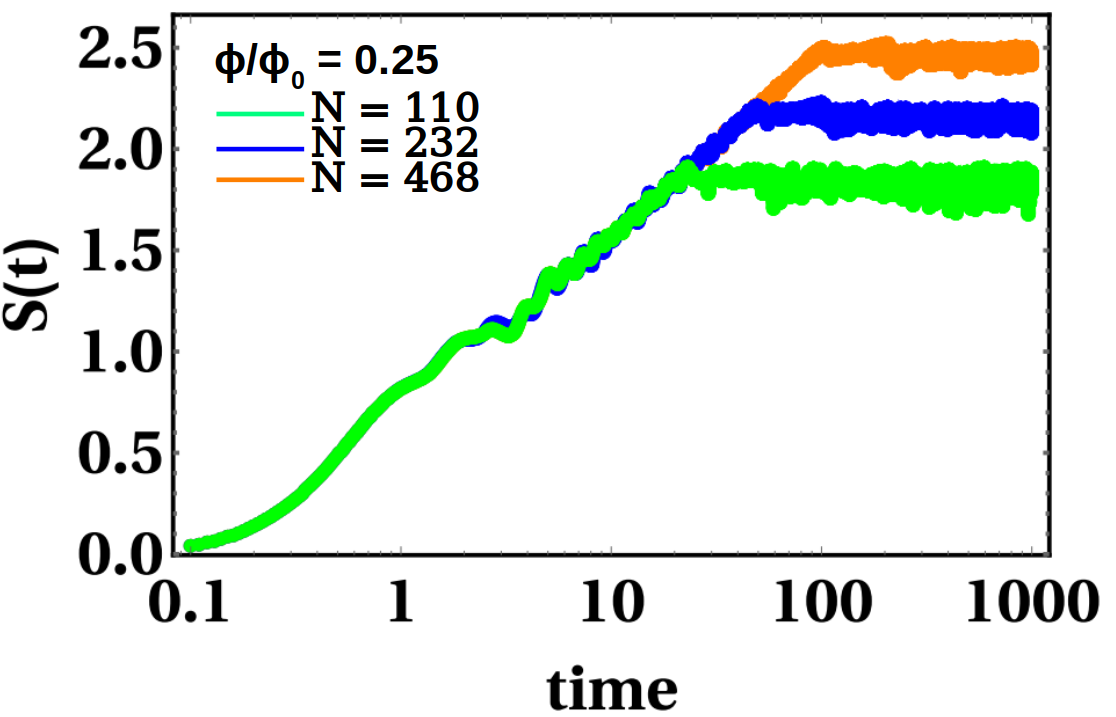}
(b)\includegraphics[width=.8\columnwidth]{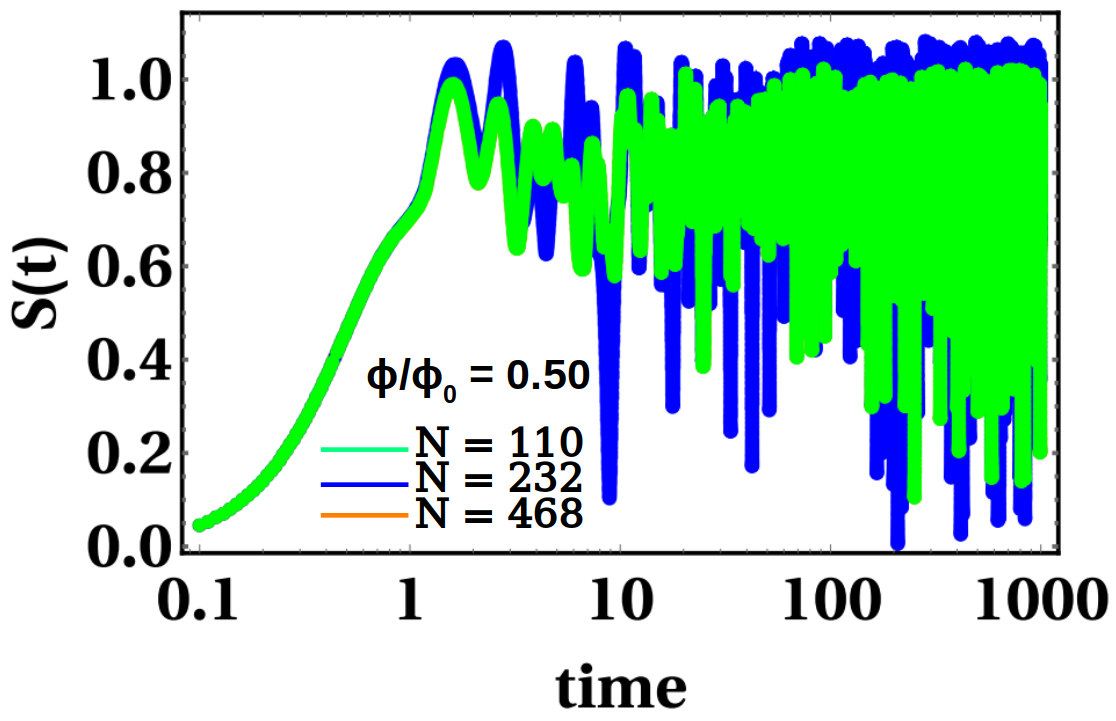}
\caption{(Color online) The information entropy is plotted against time (in a log scale) for a Fibonacci sequence of diamond-shaped network for different system size $N = 468$ (orange colored), $N = 232$ (blue colored) and $N = 110$ (green colored) with stubs when a uniform magnetic flux ($\Phi$) (a) $\Phi = \frac{1}{4}\Phi_0$ and (b) $\Phi = \frac{1}{2}\Phi_0$ is trapped in each plaquette. The orange colored plot for $N = 468$ overlaps with the two other colors, and is not visible. The hopping integrals are chosen as $L = 1$ and all on-site potentials are set as zero. The initial site where the wavepacket is released is chosen as $m = 234^{th}$, $m = 114^{th} $ and $ m = 56^{th}$ for system size $N = 468$, $N = 232$  and $N = 110$ respectively. }  
\label{entropy2}
\end{figure}

In Fig.~\ref{entropy1} we calculate $S(t)$ for the Fibonacci sequence of `dimer-stub' quantum network with two different system sizes, one is $N = 290$ (orange colored) and another is $N = 120$ (green colored). For $\lambda = 2$ $S(t)$ has a monotonic increase in time and the long time saturation value increases linearly with the system size ( $N = 120$ to $N = 290$), as shown in Fig.~\ref{entropy1}(a). This type of behaviour displayed by the information entropy with the size of the system implies that the wavepacket moves towards the boundary, ensuring that the states are perfectly delocalized. In Fig.~\ref{entropy1}(b) $S(t)$ is plotted against time when $\lambda = 3.5$. The saturation sets in only after a long enough time, and this feature is connected with the slow (sub-diffusive) decay. 

The information entropy versus time for a Fibonacci sequence of a `diamond-stub' network is shown in Fig.~\ref{entropy2} for different system sizes, viz, $N = 110$ (green colored), $N = 232$ (blue colored) and $N = 468$ (orange colored). At $\Phi = \frac{1}{4}\Phi_0$, the saturation of $S(t)$ increases in proportion with the size of system. It is apparent that before the saturation is in every case $S(t)$ has an almost monotonic increase in time. This again indicates that the wavepacket under the resonance condition of $\Phi = \frac{1}{4}\Phi_0$ exhibits a purely delocalized character (Fig.~\ref{entropy2}(a)). When each diamond plaquette is trapped by the magnetic flux $\Phi = \frac{1}{2}\Phi_0$, information entropy for different system sizes shows an overlapping of the saturation zone  which is totally independent of system size (as shown in Fig.~\ref{entropy2}(b)) as a result. This is the pinned localized phase. The wavepacket gets {\it frozen}, and is localized within a very small spatial extent around the initial point of release. 

\subsection{Inverse Participation Ratio}
As a final check over all our conclusions drawn so far,  we investigate the nature of the inverse participation ratio (IPR) when the system evolves with time. The inverse participation ratio is defined as~\cite{Mauro,pastawski},
\begin{equation}
    IPR(t) = \sum_{n} |\Psi_{n}(t)|^4
\end{equation}
Here $n$ is the atomic site index. At $t = 0$ the wavepacket is on its initial location $m$, such that only $ |\Psi_{m}(t)|^4 $ gives $1$, and at all other locations $n$ it is zero. As a result $IPR(0) = 1$. 


\begin{figure}[ht]
\centering
(a) \includegraphics[width=.75\columnwidth]{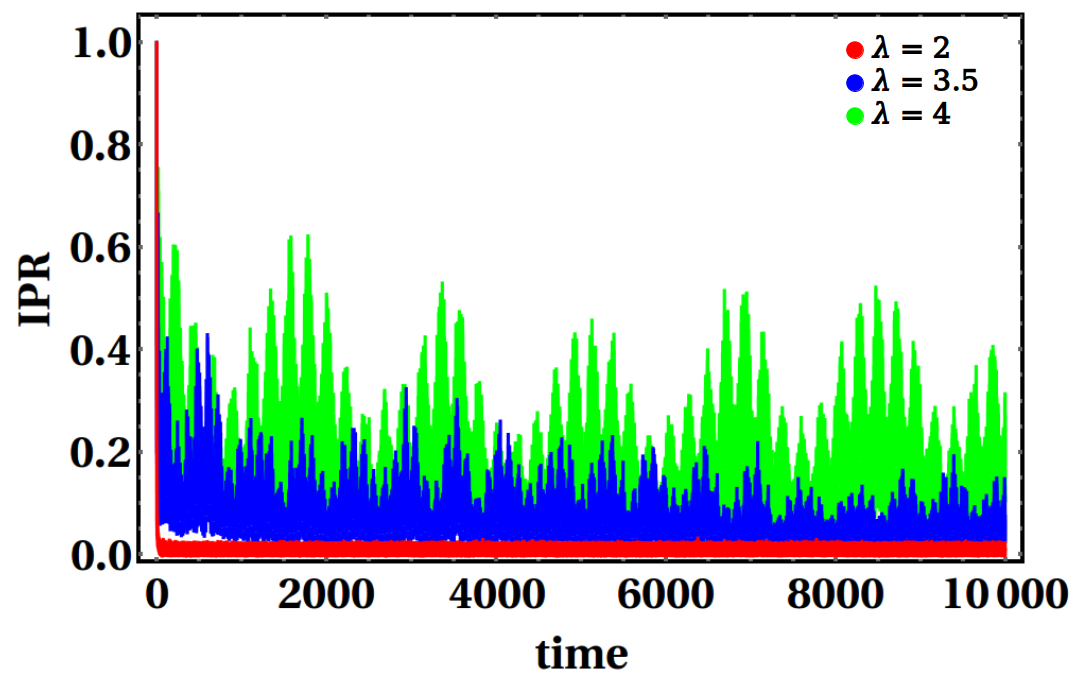}\\
(b) \includegraphics[width=.75\columnwidth]{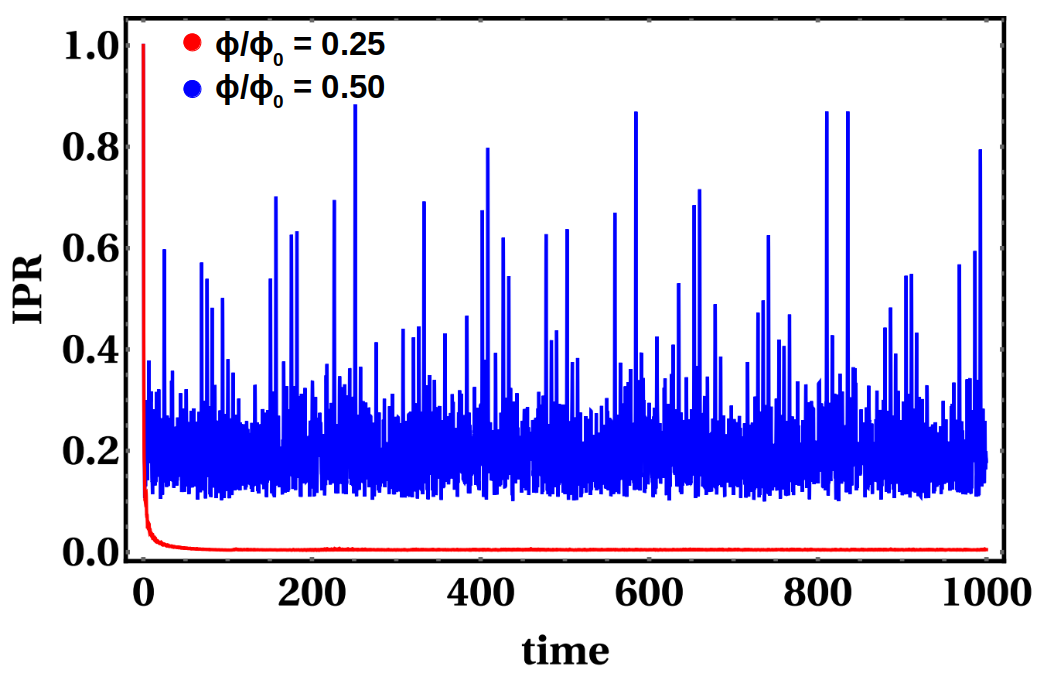}
\caption{(Color online) The inverse participation ratio is plotted against time for the Fibonacci sequence of (a) `dimer-stub' quantum network ($ 290 $ sites) with different values of the external hopping parameter $\lambda = 2$ (red colored), $3.5$ (blue colored), $4$ (green colored) and (b) for `diamond-stub' network ($468$ sites) with different values of applied magnetic field $\Phi = \frac{1}{4}\Phi_0$ (red colored), $\Phi = \frac{1}{2}\Phi_0$ (blue colored) respectively. Initially, we put the wave packet at the $141^{th}(234^{th})$ site for `dimer-stub'(`diamond-stub) systems respectively. The hopping integrals are chosen as (a) $L = 1$, $ S = \sqrt{5}$, (b) $L = 1$, and all on-site potentials are set as zero. }  
\label{ipr1}
\end{figure}

Now while delocalizing with time the wavepacket spreads out from its initial location with a non-zero probability $|\Psi_{n}(t)|^2$ on each site. This probability amplitude decreases with time. So the sum of  $|\Psi_{n}(t)|^4$ over all atomic sites tends to zero when time becomes large enough.
On the other hand, in a localized phase the wavepacket remains confined in the close proximity of its initial position. The probability $|\Psi_{n}(t)|^2$ is larger close to its initial location and at all other locations it is practically zero. Due to this type of distribution, $IPR(t)$ never decays to zero, no matter after how long the dynamics is observed.

The variation of inverse participation ratio with time for a Fibonacci sequence of `dimer-stub' quantum network is shown in Fig.~\ref{ipr1}(a) when the external hopping amplitude $\lambda$ is tuned. At resonance ($\lambda = 2$) the system is equivalent to a periodic one, and hence under time evolution, IPR(t) goes to zero (red colored). When $\lambda = 3.5$ or $4$, the dynamics represents the sub-diffusive regime the IPR shows an oscillatory behaviour between some non-zero values, slowly dropping down to zero for very long time, indicating an eventual escape of the wave packet.

A similar plot for the `diamond-stub' Fibonacci sequence is shown in Fig.~\ref{ipr1}(b) when a uniform magnetic flux is trapped in each diamond cavity. When a constant flux $\Phi = \frac{1}{4}\Phi_0$ is applied, the resultant IPR  drops to zero (red colored) with time, whereas a purely localized phase appears at  $\Phi = \frac{1}{2}\Phi_0$ and IPR is always gives a non-zero value with an oscillatory character (blue colored lines).

\section{Conclusions}
We have undertaken a detailed investigation of the quantum dynamics of a wave packet released at a point inside a couple of decorated lattices. The lattices are grown following a quasiperiodic Fibonacci sequence of bonds, along with nominal quasi-one dimensionality introduced by grafting atomic sites from one side or from both sides of the central linear backbone. In one case we have simple stub-like geometry, and in the other case, we have diamaond shaped cells that have a uniform magnetic flux trapped in them, placed along with isolated dots along the backbone. We find that under a suitable correlation among the numerical values of the hopping amplitude along the backbone and the tunnel hopping amplitude between the backbone and the stub-atom, an entire one dimensional quasiperiodic chain can be made almost transparent (with the numerical value of the end-to-end transmission coefficient becoming close to one for a long enough system), to an incoming wave packet. In the second example, a tuning of the magnetic flux does this job, providing us with a flux controlled insulator to metal transition. We first work out the possibility of such a correlation driven transition using the time independent Schr\"{o}dinger equation, and substantiate the result by studying the quatum dynamics of a wave packet released on such chains that are long but finite in length. Interestingly, the sizes of the systems chose for the dynamics already give evidence of giving full support to the conclusions drawn from the time-independent stationary state solutions. Our results and conclusions related to the commutation of the matrices and the dynamics under the resonance condition (extended eigenstates) are valid not only for the Fibonacci quasiperiodic case, but also for other extensions of the Fibonacci lattice (with identical structural units), as well as for any arrangement of the building blocks shown in this paper, including a random arrangement. However, the dynamical behaviour away from the resonance conditions should depend on the microscopic details of the lattice structures. The exponents, within the limits of numerical accuracy and the size of the system, remain very close to what we report in this work. This hints towards a new kind of {\it universality} among these microscopically different lattices. The lattices proposed in this work may be inspirational to the photonics studies, where ultrafast laser writing may be used to develop such systems. 

\section{Acknowledgements}
SB is thankful to the Government of West Bengal for the SVMCM Scholarship (WBP221657867058). AC acknowledges illuminating discussions with Muktish Acharyya.

\appendix
\section{Density of states}

To determine the average density of states (AVDOS) of the Fibonacci sequence of the `dimer-stub' or `diamond-stub' quantum network, we have used the standard Green's function technique. If $H$ is the Hamiltonian of the original system, such as in Fig.~\ref{fig}(a) or (b), then the Green's function can be written as $G = (E-H)^{-1}$. The AVDOS is given by,
\begin{equation}
    \rho_{av} (E) = Lim_{\eta \rightarrow 0} \left [-\frac{1}{N \pi} Im [ Tr~ G (E + i\eta)] \right ]
\end{equation}
where, the symbol `Tr' implies the trace of the $G$-matrix.
\section{Transmission Co-efficient}

\begin{figure}[ht]
\centering
\includegraphics[width=.95\columnwidth]{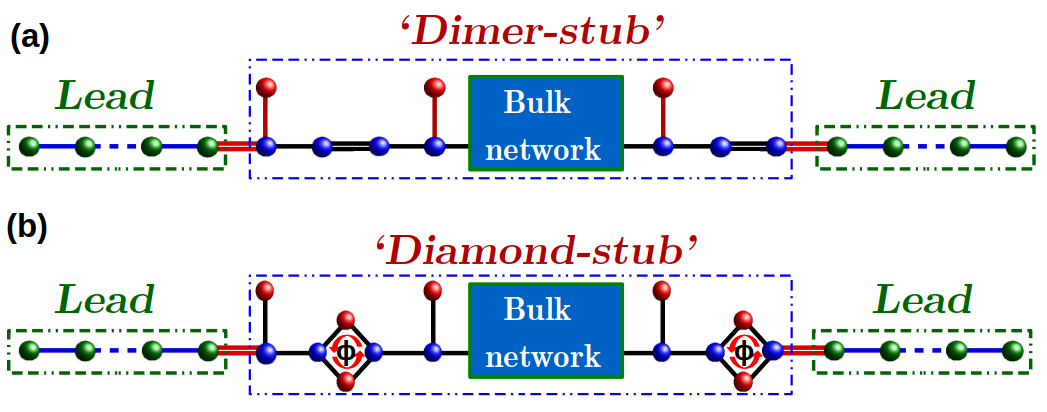}
\caption{(Color online)  schematic diagram of the Fibonacci sequence of (a) `dimer-stub' and (b) `diamond-stub' quantum network clamed in between two periodic semi-infinite leads.}  
\label{trans-fig}
\end{figure}
 To calculate the transmission coefficient of such a Fibonacci sequence of `dimer-stub' or `diamond-stub' network, we have adopted the two-terminal lead method such that the `sample' is clamped between two semi-infinite periodic `leads' (green atoms) for both geometries as shown in Fig.~\ref{trans-fig}. 
Now, with the help of the standard Green’s function technique, we can determine the transmission characteristics. The Green's function is defined as $ G = (E - H )^{-1}$, where $H$ is the Hamiltonian of the whole system, the `sample' as well as the semi-infinite leads. The Hamiltonian of the combined system can be written as, 
\begin{equation}
    H  =  H_{S} + H_{L1}+H_{L2}+H_{LS1}+H_{LS2}+H^\dagger_{LS1}+H^\dagger_{LS2}
\end{equation}
Here $H_S$ is the tight binding Hamiltonian of the sample (the system under study, as shown in Fig.~\ref{fig} (a,b) ) and  $H_{L1}(H_{L2})$ accounts for the two semi-infinite leads which are connected to the left (right) side of the system. The coupling between the lead and the system for both sides is described by the Hamiltonians  $H_{LS1}(H^\dagger_{LS1})$ and $H_{LS2}(H^\dagger_{LS2})$ respectively. The effective Green’s function of the whole combined network is given by~\cite{fisher,supriya},

\begin{equation}
    G = ( E - H_S - \Sigma_{1} - \Sigma_{2})^{-1}
\end{equation}
The Green's function of the two semi-infinite leads can be written as $G_{L1(L2)} = [E - H_{L1(L2)}]^{-1}$ and due to the connectivity of the leads with the system the `self-energy' term $\Sigma_{1(2)}$ arise, which is defined as $\Sigma_{1(2)} = H_{LS1(LS2)}G_{L1(L2)}H^\dagger_{LS1(LS2)} $. After the determination of the self-energy part, it is straightforward to obtain the coupling function $\Gamma_{1(2)}(E)$, viz.

\begin{equation}
    \Gamma_{1(2)}(E) = i \left[ \Sigma^{ret}_{1(2)}(E) - \Sigma^{adv}_{1(2)}(E)  \right]
    \label{coupling}
\end{equation}
where $\Sigma^{adv}_{1(2)}(E)$ ($\Sigma^{ret}_{1(2)}(E)$) are the advanced self-energy (retarded self-energy) and they are Hermitian conjugates of each other. Eq.~\eqref{coupling} is therefore modified as  $\Gamma_{1(2)}(E) = 2 Im \left[ \Sigma^{ret}_{1(2)}(E)  \right]$.

The  expression for the transmission coefficient can be written as~\cite{supriya}, 

\begin{equation}
    T(E) = Tr \left[ \Gamma_1(E) G^{ret} \Gamma_2(E) G^{adv}  \right]
    \label{trans}
\end{equation}


\begin{thebibliography}{99}
\bibitem{sougata1} Actually, to calculate the mean square displacement one should be using the formula, $\sigma^2(t) = \sum_n (n-<n>)^2 |\psi_n(t)|^2$, where $<n>$ is the expectation value of $n$ at a given time, instead of $m$, a specific site index. However, the dynamical behaviour does not get affected at all by this formula, as has been checked by us using extensive numerics.
\end{thebibliography}
\end{document}